%% file: PAPER_Stochastic_Loss_Reserving.tex
\documentclass[12pt]{article}
\usepackage{lmodern} 
\usepackage[english]{babel}
\usepackage{amsmath}
\usepackage{amsfonts}
\usepackage{amssymb}
\usepackage{graphicx}

\usepackage{breakcites} 

\usepackage[pdftex]{hyperref}
\hypersetup{
    plainpages=false,       
    unicode=false,          
    pdftoolbar=true,        
    pdfmenubar=true,        
    pdffitwindow=false,     
    pdfstartview={FitH},    
    pdfauthor={Andrew W.L. Fleck},    
    pdfnewwindow=true,      
    colorlinks=false,        
	linkbordercolor={1 0 0}, 	 
	citebordercolor={0 1 0}, 	
	urlbordercolor ={0 1 1},    
    linkcolor=blue,         
    citecolor=green,        
    filecolor=magenta,      
    urlcolor=cyan           
}

\usepackage{bm} 

\usepackage[affil-it]{authblk} 

\usepackage{verbatim}

\usepackage{color,soul}

\usepackage{tikz}
\usetikzlibrary{calc,decorations.markings}

\usepackage{caption}
\usepackage{subfig}
\usepackage{wrapfig,lipsum,booktabs}

\usepackage{sectsty}
\sectionfont{\centering}

\numberwithin{equation}{section}
\numberwithin{figure}{section}
\numberwithin{table}{section}

\usepackage{amsthm}
\newtheorem{theorem}{Theorem}[section]

\newtheorem{definition}{Definition}[section]

\newcommand{\inprod}[2]{ {#1}^{T}{#2} }

\newcommand{\expect}{\mathbb{E}}

\usepackage{mathbbol} 
\newcommand{\var}{\mathbb{Var}}

\newcommand{\cov}{\mathbb{Cov}}

\newcommand{\mel}[2]{ \mathcal{M}\left\lbrace {#2} \right\rbrace \left({#1}\right) }

\newcommand{\lap}[2]{ \mathcal{L}\left\lbrace {#2} \right\rbrace \left({#1}\right) }

\newcommand{\fou}[2]{ \mathcal{F}\left\lbrace {#2} \right\rbrace \left({#1}\right) }

\newcommand{\G}[1]{ \Gamma \left( {#1} \right) }

\newcommand{\sign} {{\mathrm{sign} \,}}

\usepackage[automake,toc,abbreviations]{glossaries-extra}
\input{glossaries}

\makeglossaries

\newbool{is_paper}
\booltrue{is_paper}

\begin{document}

\title{\textbf{Stochastic Loss Reserving: Dependence and Estimation}}

\author[a,c]{Andrew W.L. Fleck%
  \thanks{Email: \texttt{afleck@yorku.ca};}}
\author[a]{Edward Furman}
\author[b]{Yang Shen}
\affil[a]{\rm \small \textit{Department of Mathematics and Statistics, York University, Toronto, ON M3J 1P3, Canada}}
\affil[b]{\rm \small \textit{School of Risk and Actuarial Studies, UNSW Sydney, NSW 2052, Australia}}
\affil[c]{BP Trading and Shipping (NA Power)}
\date{}

\maketitle

\begin{quote}
\textbf{Abstract} Nowadays insurers have to account for potentially complex dependence between risks. In the field of loss reserving, there are many parametric and non-parametric models attempting to capture dependence between business lines. One common approach has been to use \textit{additive background risk models} (ABRMs) which provide rich and interpretable dependence structures via a common shock model. Unfortunately, ABRMs are often restrictive. Models that capture necessary features may have impractical to estimate parameters. For example models without a closed-form likelihood function for lack of a probability density function (e.g. some Tweedie, Stable Distributions, etc).

We apply a modification of the continuous generalised method of moments (CGMM) of \cite{carrasco2000generalization} which delivers comparable estimators to the MLE to loss reserving. We examine models such as the one proposed by \cite{avanzi2016stochastic} and a related but novel one derived from the stable family of distributions. Our CGMM method of estimation provides conventional non-Bayesian estimates in the case where MLEs are impractical.
\end{quote}

\newpage 


\input{paper2.tex}

\appendix

\clearpage

\let\part\section
\let\section\subsection
\part{Distribution Background} 
\input{DispTweedie}
\input{StableShort}

\newpage

\part{Misc tables}\label{sec: tables and data}

\input{App-F}

\clearpage
\newpage

 
\bibliographystyle{apalike}
\bibliography{York_thesis}

\end{document}

%% file: glossaries.tex
\newabbreviation{mle}{MLE}{Maximum Likelihood Estimate}

\newabbreviation{gmm}{GMM}{Generalized Method of Moments}

\newabbreviation{cgmm}{CGMM}{Continuous Generalized Method of Moments}

\newabbreviation{pdf_words}{PDF}{Probability Density Function}

\newabbreviation{cdf_words}{CDF}{Cumulative Distribution Function}

\newabbreviation{mgf_words}{MGF}{Moment Generating Function}

\newabbreviation{cf_words}{CF}{Characteristic Function}

\newabbreviation{clt}{CLT}{Central Limit Theorem}

\newabbreviation{gclt}{GCLT}{Generalized Central Limit Theorem}

\newabbreviation{edm}{EDM}{Exponential Dispersion Model}

\newabbreviation{cr}{CR}{Cramér–Rao}

\newabbreviation{VaR}{VaR}{Value at Risk}

\newabbreviation{ES}{ES}{Expected Shortfall}

\newabbreviation{TCE}{TCE}{Tail Conditional Expectation}

\newabbreviation{dy}{DY}{Development Years}

\newabbreviation{ay}{AY}{Accident Years}

\newabbreviation{cl}{CL}{Chain Ladder}

\newabbreviation{pcps}{PCPs}{Premium Calculation Principles}

\newabbreviation{capm}{CAPM}{Capital Asset Pricing Model}

\newabbreviation{apt}{APT}{Arbitrage Pricing Theory}

\newabbreviation{wipm}{WIPM}{Weighted Insurance Pricing Model}

\newabbreviation{abrm}{ABRM}{Additive Background Risk Model}

\newglossary*{symbols}{List of Symbols}

\newglossaryentry{Capital_Letter}
{
name={$X$},
sort={Prob},
type=symbols,
description={Random Variables: In general, we denote random quantities by capital letters and deterministic ones by lowercase. }
}

\newglossaryentry{Stoch_Process}
{
name={$\sigma_t$,$r_T$ or $W_t$ etc},
sort={Prob},
type=symbols,
description={Stochastic Processes: We denote families of random variables indexed by time  ($t$ or $T$) as either lower- or uppercase letters with time as the subscript.}
}

\newglossaryentry{pdf}
{
name={$f_{X}(x|\theta)$},
sort={Prob},
type=symbols,
description={Density Functions: The pdf of $X$ at a specific value $x$ given parameters $\theta$ }
}

\newglossaryentry{CDF}
{
name={$F_{X}(x)$},
sort={Prob},
type=symbols,
description={Distribution Functions: The cdf of $X$ at a specific value $x$ i.e. $P(X \leq x)$ }
}

\newglossaryentry{Expectation}
{
name={$\expect[ X ]$},
sort={Prob},
type=symbols,
description={Expected value of a random variable}
}

\newglossaryentry{Variance}
{
name={$\var[ X ]$},
sort={Prob},
type=symbols,
description={Variance of a random variable}
}

\newglossaryentry{Covariance}
{
name={$\cov[ X, Y ]$},
sort={Prob},
type=symbols,
description={Covariance of random variables $X$ and $Y$}
}

\newglossaryentry{mgf}
{
name={$M_X(\tau) = \expect[ e^{\tau X} ]$},
sort={Prob},
type=symbols,
description={Moment generating function of a random variable $X$}
}

\newglossaryentry{cf}
{
name={$\phi_X(\tau) = \expect[ e^{i \tau X} ]$},
sort={Prob},
type=symbols,
description={Characteristic function of a random variable $X$}
}

\newglossaryentry{cgf}
{
name={$K(\tau) = \log( M_X(\tau) ) $},
sort={Prob},
type=symbols,
description={Cumulant generating function of a random variable $X$}
}

\newglossaryentry{cumulant}
{
name={$\kappa_n[X] = \frac{d}{d \tau} K_{X}(0) $},
sort={Prob},
type=symbols,
description={$n$th cumulant of a random variable $X$}
}

\newglossaryentry{outprod}
{
name={$\circ$},
sort={LinA},
type=symbols,
description={Outer product: Given two vectors $\mathbf{x}$ and $\mathbf{y}$  of appropriate length, we have $\mathbf{x} \circ \mathbf{y} = \mathbf{x}  \mathbf{y}^\top $.}
}

\newglossaryentry{vector}
{
name={$\mathbf{X}$ or $\mathbf{x}$},
sort={LinA},
type=symbols,
description={Vectors: Random or deterministic vectors will be identified as boldface quantities. See Random Variables entry for more detail.}
}

\newglossaryentry{matrix}
{
name={$\mathbf{M}_{n \times m}$},
sort={LinA},
type=symbols,
description={Boldface quantities (with subscripts representing dimensions) are reserved for matrices and operators generally. For vectors, that is matrices with one column, we drop the dimension subscripts.}
}

\newglossaryentry{f_op}
{
name={For a function $f(x)$: $\mathcal{A}f$ or $(\mathcal{A}f)(x)$},
sort={LinA},
type=symbols,
description={Cursive capitals are reserved for integral operators.}
}

\newglossaryentry{transpose}
{
name={$\mathbf{M}_{n \times m}^\top $},
sort={LinA},
type=symbols,
description={Matrix Transpose}
}

\newglossaryentry{mellin}
{
name={$\mel{s}{f(x)}$},
sort={IntTrans},
type=symbols,
description={Mellin transform of $f$ given by $\int_0^{\infty} x^{s-1} f(x) dx $}
}

\newglossaryentry{laplace}
{
name={$\lap{s}{f(x)}$},
sort={IntTrans},
type=symbols,
description={Laplace transform of $f$ given by $\int_0^{\infty} e^{-sx} f(x) dx $}
}

\newglossaryentry{fourier}
{
name={$\fou{s}{f(x)}$},
sort={IntTrans},
type=symbols,
description={Fourier transform of $f$ given by $\int_{-\infty}^{\infty} e^{-isx} f(x) dx $}
}

\newglossaryentry{premium}
{
name={$\pi[X]$},
sort={ActSci},
type=symbols,
description={Premium Principle: The premium paid (in currency) for a risk represented by a random variable $X$ }
}

\newglossaryentry{risk_measure}
{
name={$H[X]$},
sort={ActSci},
type=symbols,
description={Risk Measure: A functional over a suitable set of loss random variables mapping losses to \textit{risk capital}  }
}

%% file: paper2.tex

\section{Introduction}

Highly regulated and vitally necessary, the loss reserve is typically the largest liability on an insurance company's balance sheet. Proper estimation of future claims is therefore paramount for financial stability. In fact, with the introduction of regimes like Solvency II actuaries are now sometimes required not just to estimate reserves but model potential shortfalls and risks of insolvency. This makes a \textit{stochastic} model of loss reserves necessary (see \cite{wuthrich2008stochastic}  or \cite{FROHLICH2018130} and references therein). This is complicated by the fact that delays between an incurred claim and proper reporting can take some time, often years. There may also be ongoing or renewed liability at a later date for many reasons, such as legal proceedings and lengthy investigations. In the recent past, there have been several high profile examples of these kinds of ``tort liabilities'' such as asbestos and other environmental pollutants (\cite{carmean1995environmental} and \cite{madigan2003reserving}). Newer concerns such as the health risks of engineered materials may present similar issues (\cite{mcalea2016engineered}). Complicating estimation even further is the potential dependence among claims between business lines.  One example that could induce such a dependence may be industry-specific inflationary trends. Medical costs can often rise faster than economy-wide price levels; accident business lines especially may need to incorporate this into their reserves. Similarly, auto repair techniques may incur increased costs for both commercial and personal lines. Such a dependence can represent a potential for diversification or an increased risk to the insurer (see e.g.\ \cite{de2012modeling}).

Given the importance of proper loss reserving it is unsurprising that there are as many forms of reserve estimation as techniques in statistics (see \cite{wuthrich2008stochastic} and references therein). In this paper we focus on the model popularized by \cite{merz2013dependence} and model claims parametrically and dependence ``cell-wise" across business lines. Within this framework, it is popular to model severity and dependence separately via copulas (e.g.\ \cite{zhang2013predicting}). We instead take a multivariate modelling approach as in \cite{avanzi2016stochastic}. The benefits and drawbacks of these two approaches (copulas vs multivariate models) are essentially the same as in traditional statistics. Copulas provide a great deal of model flexibility but at the cost of increased numbers of parameters and decreased interpretability. Multivariate models are much more parsimonious but restrict the available marginals. 

In order to negotiate this trade-off we construct incremental loss models across business lines via an \gls{abrm}. ABRMs provide an easily interpretable and flexible dependence through the use of a common shock structure across business lines. The technique can be easily extended to a variety of marginal distributions leading to many possible multivariate models. By way of example, this \ifbool{is_paper}{paper}{chapter} makes use of the multivariate gamma and Tweedie model of, respectively, \cite{furman2005risk} and \cite{furman2010multivariate} as in \cite{avanzi2016stochastic} as well as introduces a particular multivariate Stable distribution. The idea of additive risk models is not new in economics and finance. The most famous examples are of course the \gls{capm} \cite{fama2004capital} and \gls{apt} \cite{ross2013arbitrage}. More recently the potential for applications in insurance -- especially enterprise risk management -- has been explored (see e.g.\ \cite{furman2018weighted} and \cite{zhou2018approximation}). Other ways to introduce dependence in loss reserves are the Multiplicative Background Risk Models (MBRMs) (e.g., \cite{furman2021multiplicative}, \cite{asimit2016background}, \cite{semenikhine2018multiplicative} and \cite{marri2022risk}), minimum-based background risk models (e.g., \cite{asimit2010multivariate} and \cite{pai2020livestock}), and background risk models that allow for multiple types of risk factors (e.g., \cite{su2017multiple} and \cite{su2017multiple1}). 

The main contribution of our work is not just applying ABRMs but also model estimation. Useful loss models frequently lack a closed form or computationally simple \gls{pdf_words}, making classical estimation difficult (e.g.\ compound Poisson, NIG in the Tweedie case or most non-normal stables). The small sample sizes and many parameters in reserve models naturally lead many to rely on a Bayesian analysis (see for example \cite{zhang2013predicting}). There have been attempts to study estimation in the multivariate Tweedie using the method of moments (\cite{alai2016multivariate}) but for the reasons already stated this seems inappropriate. \gls{cgmm} of \cite{carrasco2000generalization} offers a hope of success where the basic GMM may fail. Incorporating an infinite number of moment conditions makes the CGMM maximally statistically efficient. In this work we outline a novel use of the CGMM that is relatively computationally inexpensive, especially for ``larger" multivariate models. By using the CGMM in conjunction with ABRMs we open up the practical application of a variety of models without the need to resort to highly uncertain Bayesian estimates. 

This \ifbool{is_paper}{paper}{chapter} is organized as follows. In Section \ref{sec: LossRes}, we give a more detailed account of cell-wise loss modelling and basic estimation. A discussion of what makes a useful model and the introduction of our Tweedie and Stable ABRM examples takes place in Section \ref{sec: ABRMs}. The CGMM and our novel approach are outlined in Section \ref{sec: CGMM}. Finally, some simulation results and an illustration using real data are given in Sections \ref{sec: LossSims} and \ref{sec: RealLoss}.

\section{Parametric Loss-Reserving Models}\label{sec: LossRes}

In this section we briefly review parametric loss-reserving models. Let us consider the following situation. We are in the $m$th  accident year since writing a particular non-life policy. We have been able to observe incremental claims $x_{i,j}$ for each of $i=1,\ldots,n$ \gls{ay}\footnote{Year losses were incurred but not necessarily paid.} and $j=1,\ldots,m-i+1$ \gls{dy}. We assume that all these are individual samples from random variables $X_{i,j}$ that are stochastically independent. We add some necessary structure by enforcing shared parameters in a typical way (see \cite{wuthrich2008stochastic}):

\begin{align}
\expect [X_{i,j}] = \mu_{i,j} = \eta_i \nu_j \quad {\rm and} \quad \var[X_{i,j}] = \sigma_{i,j} = w_{i,j} \gamma_j 
\end{align}

\noindent where $w_{i,j}$ is an appropriate weight and $\gamma_j$ a scale parameter. For some $X_{i,j}$ with a \gls{pdf_words} of the form \glssymbol{pdf}[$f_{X_{i,j}}( x_{i,j} |\mu_{i,j} , \sigma_{i,j})$], such as Tweedie, Stable, and so on, we can easily construct a \gls{mle} for the model parameters:
\begin{align}\label{eq:MLE}
\lbrace ( \hat{\eta}_i, \hat{\nu}_j, \hat{\gamma}_j) \rbrace_{i,j}^m = \underset{\eta_i, \nu_j, \gamma_j}{\mathrm{argmin}} \left\{ \prod\limits_{i,j}^{i+j<m+1} f_{X_{i,j}}( x_{i,j} |\mu_{i,j} , \sigma_{i,j}) \vert_{\mu_{i,j} =\eta_i \nu_j , \sigma_{i,j} = w_{i,j} \gamma_j }  \right\} .
\end{align}

\begin{figure}[h!]
  \centering
  \includegraphics[scale=0.5]{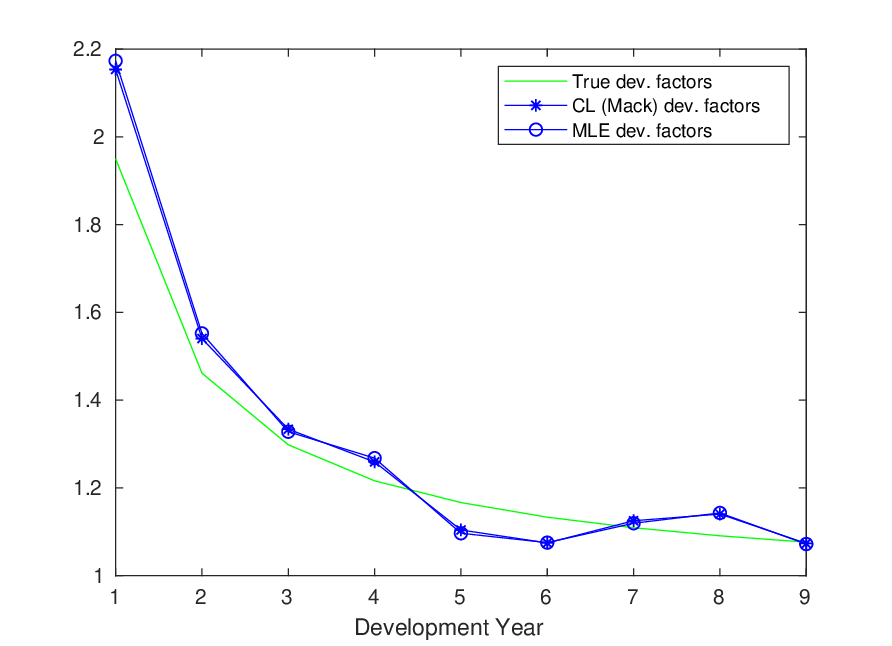}
  \caption{Comparison of the true, MLE- and CL-estimated development factors in a Gamma-distributed development triangle, specifically $p=2$ with $\gamma=\gamma_j= 0.2$ , $\eta_i = \eta =5$ and $\nu$ ranging over 1 to 0.55}
 \label{fig: Tweedie Ex}
\end{figure}

In fact, for Tweedie-distributed incremental losses, estimators of the form (\ref{eq:MLE}) are numerically similar to the well-known \gls{cl} estimators of \cite{mack1993distribution} (see \cite{mack1991simple} or especially \cite{taylor2007chain} for more details). Notably, for a Tweedie power parameter of $p=1$ (the overdispersed Poisson) the correspondence is exact. For values ``close" to $p=1$, these estimates are \textit{very} similar conditional on the scale. For example, if we consider a Gamma model of the form
\begin{equation}\label{exp:Gamma model}
X_{i,j} \sim \text{Gamma} \left( \frac{1}{\gamma_j} , \eta_i \nu_j \gamma_j \right) \sim Tw_2 ( \eta_i \nu_j , \gamma_j) 
\end{equation} 

\noindent where $Tw_p(\mu, \sigma^2)$ denotes the reproductive Tweedie density (see \ref{app: Tweedie} for definitions and details), we can simulate a Gamma-distributed loss development triangle and compare the \gls{mle}-derived development factors to those from the \gls{cl} estimates in Figure \ref{fig: Tweedie Ex}. It can be seen that the \gls{cl} and \gls{mle} estimates are virtually identical and both of them are fairly good estimates for the true model. 

\begin{figure}[ht]
\hspace*{-1.525cm}
  \centering
  \subfloat[]
  {
  \includegraphics[scale=0.4]{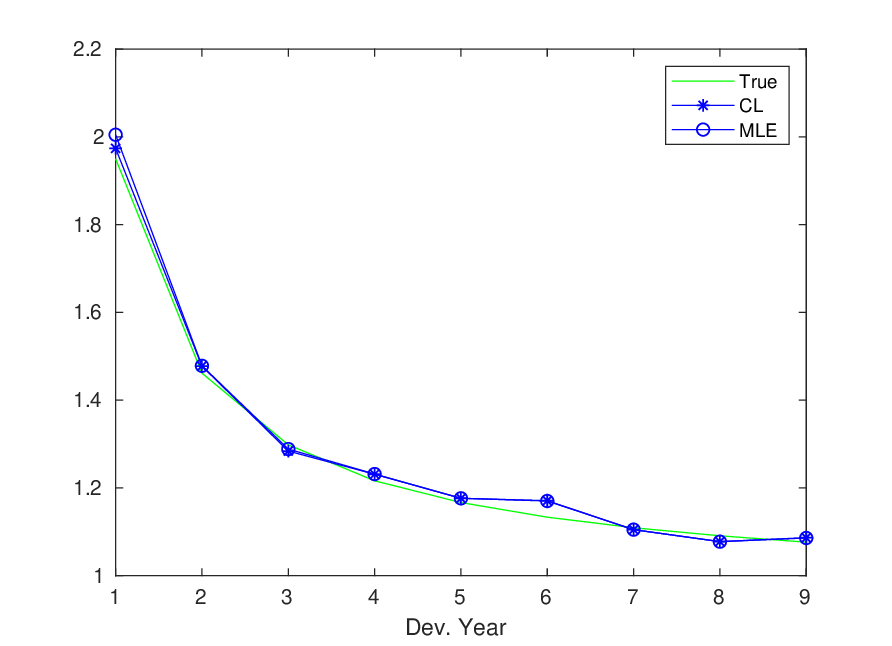}\label{fig: Stable Ex1}
  }
  \subfloat[]
  {
  \includegraphics[scale=0.4]{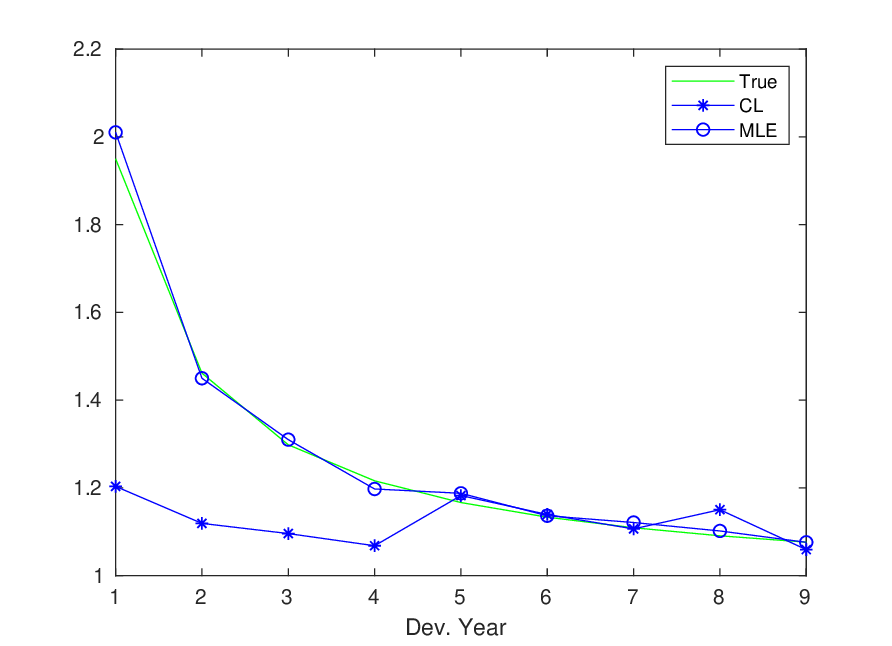}\label{fig: Stable Ex2}
  }
  \subfloat[]
  {
  \includegraphics[scale=0.4]{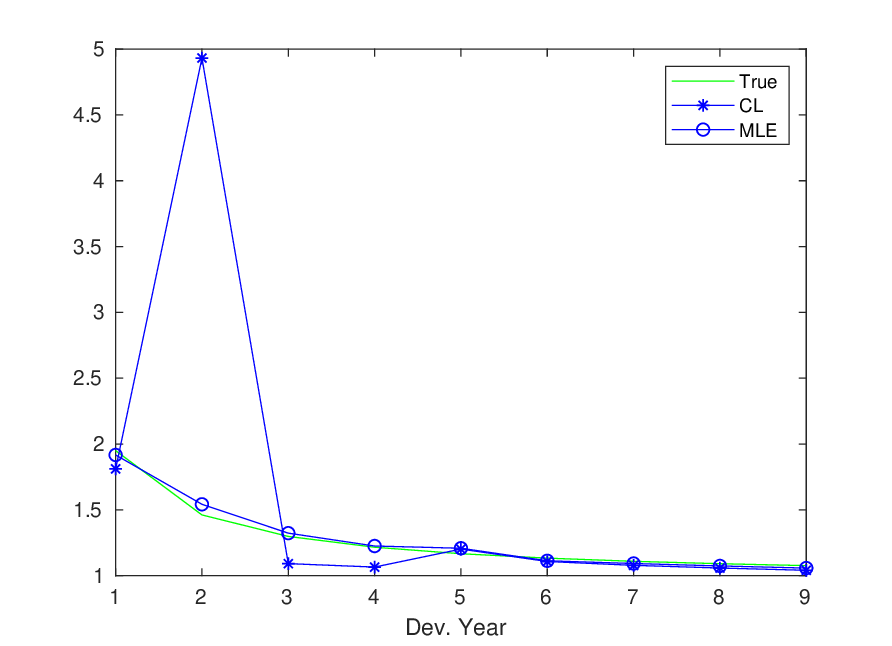}\label{fig: Stable Ex3}
  }
  \caption{Examples of draws from the same development triangle ($\eta =5$ and $\nu= 1 - 0.55$) featuring Stable-distributed losses with tail parameter $\alpha=1.8$}\label{fig: Stable Exs}
\end{figure}

While the Tweedie class of distributions is quite large, there are some losses for which it is inappropriate. For example, fire and automobile insurance coverage frequently exhibit heavy Pareto-style tails, suggesting an infinite or undefined variance (see \cite{seal1980survival}). The behaviour of such heavy-tailed distributions is qualitatively different from the typical Tweedie models near $p=1$ (or thin-tailed models generally). One extremely useful and well-motivated model is the Stable distribution. If we repeat our experiment from Figure \ref{fig: Tweedie Ex} with a stable loss model we can see in Figure \ref{fig: Stable Exs} how the \gls{cl} estimates quickly break down. 

In the stable case, many draws would exhibit fairly typical behaviour (see Figure \ref{fig: Stable Ex1}). This is not too surprising as the CL estimates are unbiased provided the mean exists. In the typical case, the sample is mostly represented by the fairly well-behaved centre of the distribution. However, successive draws can reveal that even one or two losses reported out on the tail of the distribution can quickly contaminate the CL estimates, rendering them useless (refer to Figure \ref{fig: Stable Ex2} and \ref{fig: Stable Ex3}). It is worth pointing out that in Figure \ref{fig: Stable Ex2} a major loss early in the development pattern had the unexpected effect of \textit{underestimating} the development factors. This may not be so obvious in a less marked example when a sample from the tail is creating similar issues. Though the examples chosen are extreme by design, it is easy to see that blind application of non-parametric estimates is ill-advised.

To fully appreciate the implications of heavy tails, we consider that in Figure \ref{fig: Stable Ex3} a single cell incurred about 50\% of all losses in the triangle. In multiple business lines sharing systematic risk, this can be even more consequential. A single draw from a significantly heavy-tailed risk shared across a portfolio could potentially be greater than the reserve estimated in a thin-tailed model. Such counterintuitive behaviour for heavy tails cannot be ignored. 

While a dramatic level of risk exists in heavy-tailed losses, there is hope of overcoming this issue. Unsurprisingly, the MLE estimates in Figure \ref{fig: Stable Exs} are adequate, but this is not without major qualification. First, in order to compute the MLEs one needs a PDF. In the stable case (excepting the Normal distribution), there is no closed-form PDF with finite variance. Calculating the PDF requires the numerical inversion of the \gls{cf_words} or the evaluation of a truncated infinite series to some precision. This numerical quadrature can be very expensive and in the case of multiple business lines is not practical. Additionally, the level of precision required is very high. In the MLEs we studied, we found that the scale parameter was often poorly reported. Furthermore, the method is very sensitive to the choice of initial points used in the optimization. In the case of Figure \ref{fig: Stable Ex2}, a few attempts had to be made before producing the results shown. 

In this \ifbool{is_paper}{paper}{chapter} we motivate the use of stable loss models. We also extend them to the case of multiple lines of business with a stable \gls{abrm}. The ABRM has a flexible and easily interpretable dependence structure modelling cell-wise dependence with a multivariate stable distribution. In order to deal with the aforementioned estimation challenges, we make use of the \gls{cgmm} of \cite{carrasco2000generalization}. These estimates are comparable to MLEs in a more typical case with an order of magnitude fewer calculations. This is achieved by constructing estimators from the \gls{cf_words} directly as opposed to reconstructing a PDF. For example, in the examples above with identical optimization parameters, the CGMM was about thirty times faster than the MLE (60 vs 2000 seconds on a standard IBM ThinkPad). We show that the CGMM makes the multivariate estimation of multiple business lines a practical reality for stable ABRMs.

\section{An Additive Background Risk Model} \label{sec: ABRMs}

While a single loss development triangle can be used to estimate the required loss reserve facing that line of business, an insurance company typically manages a portfolio of several lines. Given the importance of reserve estimates, dependence between reserves must be modelled. To that end, we identify a few desirable properties of any cell-wise dependent model:

\begin{itemize}
\item Marginal flexibility: Any model must capture a large class of possible distributions.

\item Closure under marginal convolution: To model sub-portfolios accurately, we must be able to add incremental cells together easily.

\item Model confidence: Our model should exist in the limit of a set of reasonable stochastic models, as for example the Normal arises from the \gls{clt}. This allows us to feel confident applying the model in general situations. 

\item Simple and flexible dependence: Any dependence must be transparent and easily interpretable. 

\end{itemize}

\noindent To establish our last desired property we introduce an ABRM. The assumptions behind this model are simple. For the $k$th line of business (LoB) of a total of $n$, we assume a cell-wise model of the form 

\begin{equation}\label{eq: ABRM cell}
X_{i,j}^{(k)} = a_{i,j}^{(k)} Y_{i,j}^{(k)} + b_{i,j} Z_{i,j}  .
\end{equation}

That is, we have an independent idiosyncratic component $Y_{i,j}^{(k)}$ and a ``common shock'' component $Z_{i,j}$ across business lines. If we denote 

$$\mathbf{a}_{i,j}=(a_{i,j}^{(1)}, ... , a_{i,j}^{(n)})' \quad\text{and}\quad \mathbf{b}_{i,j}=(b_{i,j}^{(1)}, ... ,b_{i,j}^{(n)})'$$ 

\noindent then we can arrange the same cells across business lines into a vector form: 

\begin{equation}\label{eq: ABRM vector}
\mathbf{X}_{i,j} = \mathbf{a}_{i,j} \circ \mathbf{Y_{i,j}} + \mathbf{b}_{i,j} Z_{i,j}  
\end{equation}

\noindent where \glssymbol{outprod}[``$\circ$"] denotes the outer product. In this way, we can reduce dependence to the $\mathbf{b}_{i,j}$ parameters alone. In fact, we can show that for some models (e.g.\ Normal) this is related to the typical Pearson correlation structure. 

The density of such a model can be found by integrating the univariate PDFs \`{a} la equation (2.9) in \cite{avanzi2016stochastic} or equation (6.1) of \cite{furman2010multivariate}. As the PDFs may not always exist, we are more concerned with the \gls{cf_words}. Letting $\mathbf{t} = (t^{(1)},...,t^{(n)})'$, we compute the multivariate CF in terms of the \glssymbol{cf}[univariate CFs]:

\begin{align}
\phi_{\mathbf{X}_{i,j}} ( \mathbf{t} ) &= \expect \left[ \exp\left\lbrace i\inprod{\mathbf{t}}{ \mathbf{X}_{i,j} }\right\rbrace \right] \\
&= \expect \left[ \exp\left\lbrace \sum_{k=1}^n i t^{(k)} a_{i,j}^{(k)} Y_{i,j}^{(k)} + i\inprod{ \mathbf{t}}{ \mathbf{b}_{i,j} }Z_{i,j} \right\rbrace \right] \\
&= \prod_{k=1}^n \phi_{ Y_{i,j}^{(k)} }(t^{(k)} a_{i,j}^{(k)}) \phi_{ Z_{i,j} } (\inprod{\mathbf{t}}{\mathbf{b}_{i,j} } )  \label{multiCF}
\end{align}

\noindent where $\inprod{ \mathbf{x} }{ \mathbf{y} }$ denotes the inner product. We can see how the value $\inprod{\mathbf{t}}{\mathbf{b}_{i,j} }$ will control the dependence structure of the distribution. We now move on to some specific examples to illustrate why this representation may be more appropriate. 

\subsection{Example 1: A Multivariate Tweedie Approach}

In our first example we discuss the multivariate Tweedie's desirable properties. It satisfies the requirements of marginal flexibility and closure under marginals. As discussed, special cases of the Tweedie are the Poisson and Gamma distributions; a more exhaustive list would include the compound Gamma, Normal, and Normal Inverse Gaussian distributions. Additionally, there is a Tweedie class generated by Stable distributions. Being an exponential dispersion model, the multivariate Tweedie has many desirable properties for inference, computation, and interpretation. 

We can construct the Tweedie ABRM model in the form of Eq.\ (\ref{eq: ABRM cell}), as follows:
\begin{align}\label{tweedie rep model}
Y_{i,j}^{(k)} &\sim Tw_p( \eta_i^{(k)} \nu_j^{(k)} , \gamma^{(k)} ) , \\
Z_{i,j} &\sim Tw_p(\alpha, \beta ) 
\end{align}
where the additional model parameters are 
\begin{align}\label{tweedie rep parm}
a_{i,j}^{(k)} &= 1 ,
 \\
b_{i,j}^{(k)} &=\left( \frac{\alpha}{\eta_i^{(k)} \nu_j^{(k)}} \right)^{1-p}\frac{\gamma^{(k)}}{\beta} .
\end{align}
This gives us a cell-wise distribution of
\begin{equation}\label{tweedie X}
X_{i,j}^{(k)} = Tw_p\left( \eta_i^{(k)} \nu_j^{(k)}\left[ \left( \frac{\alpha}{\eta_i^{(k)} \nu_j^{(k)}} \right)^{2-p}\frac{\gamma^{(k)}}{\beta} + 1 \right] ,\gamma^{(k)} \left[ \left( \frac{\alpha}{\eta_i^{(k)} \nu_j^{(k)}} \right)^{2-p}\frac{\gamma^{(k)}}{\beta} + 1 \right] \right) .
\end{equation}

For various values of the Tweedie power parameter $p$, we lack a closed-form PDF. We can however easily construct the univariate CF via the cumulant function of the Tweedie distribution (see sections \eqref{app: EDM} and \eqref{app: Tweedie} for details). 
If $Y\sim Tw_p (\mu, \sigma^2)$, then 
\begin{align*}
\expect[e^{itY}]= \exp\left\{ \frac{1}{\sigma^2} \left[\kappa_p (g(\mu) + it \sigma^2 )-\kappa_p (g(\mu)) \right]\right\} 
\end{align*}
where $\kappa_p(\cdot)$ is given by equation \ref{eq: kp}. 

The universality properties of the Tweedie are perhaps what makes it most applicable to insurance loss modelling. Tweedie distributions are the only EDMs that are closed under rescaling. Take any \gls{edm} whose variance is asymptotically a function of the mean such that $\var[Y] \sim C \cdot \expect[Y]^p$. This model ``converges'' in some sense with rescaling to a Tweedie distribution (\cite{jorgensen1987exponential}). 

Abusing notation slightly, suppose $ED(\mu, \sigma)$ is a random variable with an exponential dispersion distribution function (\ref{eq:REDM}). Reading the following equality in the sense of distribution, if
\begin{equation*}
cED(\mu, \sigma) = ED( c\mu, c^{2-p} \sigma^2) 
\end{equation*}
then $ED(\mu, \sigma)$ is Tweedie. EDMs converge under rescaling to Tweedies, in much the same way that Stable variables remain stable under addition and the limit of normalized sums. For any $\mu >0$ and $\sigma^{2}>0 $, suppose $\var[Y] = \sigma^2 V(\mu)$ (\ref{eq: var fun}); if $V(\mu) \sim c_{0} \mu^p $ as $c\rightarrow 0$ (resp.\ $\infty$), then we have 
\begin{align*}
\frac{ED(c\mu ,\sigma ^{2}c^{2-p})}{c} \rightarrow Tw_{p}(\mu ,c_{0}\sigma ^{2}) 
\end{align*}
in distribution as $c\rightarrow 0$ (resp.\ $\infty$). Consider some hypothetical losses incurred according to an EDM with $s, \mu$ and $s\sigma$ where $s$ is the scale (\$1,000's, \$1M's etc.) as long as the variance remains constant at scale (i.e.\ $p=0$). In this case, 
\begin{align*}
\frac{ED(s\mu ,(s\sigma)^{2})}{s} \rightarrow Tw_{2}(\mu ,s_{0}) =\mathcal{N}(\mu,s_{0}) .
\end{align*}

Another example would be for discrete data. Imagine that we are modelling the number of claims $N$ in some risk model with distribution $N \sim ED(\mu,1)$, where $V(\mu) \sim \mu$. Then
\begin{align*}
\frac{ED(c\mu ,c)}{c} \rightarrow Tw_{1}(\mu,1) = Poi(\mu) .
\end{align*}

A trivial example of this would be if claims were binomially distributed and the number of claims were aggregated or rescaled. The compound Poisson--Gamma is a Tweedie distribution (a compound distribution of two other Tweedies) that may be obtained using this mechanism, which may explain its popularity as a loss model. 

\subsection{Example 2: A Multivariate Stable Approach}\label{subsec: stable abrm}

We now consider the most ``natural'' model that would meet all our required criteria: the multivariate Normal distribution. As the basin of attraction in the Central Limit Theorem, it has the required closure under convolutions and an interpretable linear dependence structure. In fact, for Tweedie power parameter $p=0$ we recover this model. 

Since we lack marginal flexibility, the symmetric nature of the Normal makes it awkward for insurance applications. We can however generalize the Normal to the \textit{Stable} family of distributions. Furthermore, we can consider the class of totally skewed Stables for a more realistic application to insurance. Stables exist as the basin of attraction of sums of heavy-tailed random variables. Thus, given the data compiled by \cite{eaton1971extreme} and \cite{embrechts2013modelling}, Stable models are a good candidate for heavy-tailed insurance portfolios, which \textit{cannot} be captured by the Tweedie ARBM. \ifbool{is_paper}{}{Indeed, we will expand on this model for insurance losses in Section \ref{SEC: Model}. }

Proceeding in the same way as before, 
\begin{align}\label{stable model}
Y_{i,j}^{(k)} &\sim S_\alpha( \eta_i^{(k)} \nu_j^{(k)} , \gamma^{(k)}, 1 ) , \\
Z_{i,j} &\sim S_\alpha(\mu, \sigma, 1 )  ,
\end{align}
we can set
\begin{equation}\label{stable param}
a_{i,j}^{(k)} =b_{i,j}^{(k)} =1 .
\end{equation}

In fact we can show that for these parameter values $\mathbf{X_{i,j}}$ is a true multivariate Stable vector with marginals 
\begin{equation}\label{stable X}
X_{i,j}^{(k)} = S_\alpha(\eta_i^{(k)} \nu_j^{(k)}+\mu, ( (\gamma^{(k)})^{\alpha} + \sigma^{\alpha} )^{\frac{1}{\alpha}} ,1) ,
\end{equation}
where the mean, if it exists, is $\mu$, and we specify scale $\sigma$ and importantly the skewness parameter $\beta$. We consider the special case of maximum skewness i.e. $\beta=1$ for the rest of the \ifbool{is_paper}{paper}{chapter} as it is the most appropriate for loss modelling. The parameter $\alpha$ controls the heavy-tailedness of the distribution. 

A Stable model for losses could arise simply through aggregation of smaller losses. Whenever there is an incremental loss, it can be written as the sum of smaller i.i.d.\ losses $L_i$: 
\begin{equation}\label{Stable_Scheme}
X_n = \frac{L_1 + \cdots + L_n}{p_n} - q_n .
\end{equation}
Then by the Generalized Central Limit Theorem (Theorem \ref{thm: GCLT}), we have that $f_{X_n} \rightarrow f_X$ weakly where $f_X$ is a standardized Stable distribution:
\begin{align*}
X \xrightarrow[]{dist.} S_\alpha(1,\beta,0) .
\end{align*}

In both the ABRMs just discussed, we often lack a closed-form PDF of the kind we had in the earlier Gamma model (Eq.\ (\ref{exp:Gamma model})). In the Stable case, for instance, we only have a closed-form PDF for tail parameters $\alpha=0,5,1,2$. Unlike our simple example earlier we also have many more parameters across $n$ business lines. This further compounds our computational issues. 

And yet we have shown such models are desirable. This raises the question: Is there a way of estimating reserves given the observed part of a loss triangle using only what we are guaranteed to have, for example the CF given by Eq.\ (\ref{multiCF})? Also, what sacrifices in terms of efficiency would this imply, if any? We explore these issues in the following section. 

\section{Estimation via Continuous Generalized Method of Moments}\label{sec: CGMM}

\subsection{Motivation}

As seen in Section \ref{sec: LossRes}, given the cell-wise PDF of our claims it is fairly easy to construct maximum likelihood estimators for our model. Unfortunately, as pointed out in the previous section, the ABRMs can lack such a PDF for many parameter values. Indeed, this may make even simple cases (single loss triangles, not too many AY/DYs) computationally expensive. As a point of comparison, one estimation from Table \ref{tab:Stable 1 LoB table} in Section \ref{sec: 5.1} with identical optimization parameters is about thirty times faster using the continuous generalized method of moments (CGMM) objective than with the MLE objective. This is due to the fact that every evaluation of the PDF/likelihood requires a numerical quadrature of a characteristic or moment function. In multiple lines of business, this is made even worse by the need to add another quadrature for a convolution of univariate PDFs.

In such a case, it may be preferable to seek out alternatives to likelihood estimation. This is the approach adopted by \cite{avanzi2016stochastic} in relation to \cite{furman2010multivariate}, employing a Markov chain Monte Carlo approach to the completion of a Bayesian analysis. While  experience rating can doubtless help the estimation procedure along, we would hope this is not the only recourse available, especially given the often counterintuitive hidden risks of heavy-tailed models specified as in Eq.\ (\ref{stable X}). 

The only other method of inference introduced in the multivariate Tweedie case is the method of moments (see \cite{alai2016multivariate}). There has also been much interest in using a method of moments-style estimator for Stable models, but with the characteristic function as a moment condition (see e.g.\ \cite{bee2018characteristic} or \cite{koutrouvelis1980regression}). However, this is inappropriate for loss reserving due to a lack of large enough sample sizes. That said, there is a generalization we may use. 

First, let us review the classical \gls{gmm} introduced by \cite{hansen1982large}. The GMM has become overwhelmingly popular in econometrics. This success is in part due to the fairly arbitrary moment conditions required and the lack of distributional assumptions. The GMM can for example handle complex nonlinear regressions involving tricky economic concepts such as endogenous variables. 

Unfortunately, the wide applicability of the GMM comes at the cost of statistical efficiency. Because they ``throw away'' the excess information of perfect specification and introduce ad hoc moment conditions, GMM estimators are widely viewed as less efficient than their more onerous MLE equivalents\footnote{There are recent optimal efficiency results (see \cite{ackerberg2014asymptotic} for two-step semi-parametric models) but this is unsurprising as we will shortly see.}. This is one reason why we should be skeptical about applying GMM methods to the small samples in loss reserving. That said, it is not hard to show that if the moment conditions are the score of the correctly specified distribution, we actually recover the MLE. To see this, we consider a sample of $\lbrace \mathbf{x}_1,..,\mathbf{x}_l \rbrace$ i.i.d.\ realizations from some random variable $\mathbf{X}$ and for $r \in \{1,...,l\}$ specify moment conditions of the form
\begin{equation}\label{ind moment cond}
\expect[ \mathbf{g}(\theta : \mathbf{x}_r) ]=0 , 1 \leq r \leq l
\end{equation}
where $\mathbf{g}(\theta: \mathbf{x}) = (g_1(\theta :\mathbf{x}),g_2(\theta :\mathbf{x}),...,g_{d_g}(\theta :\mathbf{x}))$ is a vector-valued function of some model parameters $\theta$, $dim(\mathbf{x})=d_x$, $dim(\mathbf{g})=d_g$, $dim(\theta)=d_{\theta}$ and $d_g>d_\theta$; thus Eq.\ (\ref{ind moment cond}) is \textit{overdetermined}\footnote{Note abuse of notation: technically we should denote $\theta$ by $\boldsymbol{\theta}$, but as a set of parameters we want to set it aside for succinctness and to avoid confusion with data and other vectors.}. We can take a sample average to approximate (\ref{ind moment cond}):
\begin{equation}\label{eq: pop moment cond}
\mathbf{g}_l(\theta) = \frac{1}{l} \sum_{r=1}^l \mathbf{g}(\theta : \mathbf{x}_r) .
\end{equation}
For some \glssymbol{matrix}[positive definite matrix $\mathbf{W}_{d_g \times d_g}$], we can define an inner product and corresponding norm and minimize the norm of Eq.\ (\ref{eq: pop moment cond}) to arrive at the GMM objective function. The GMM estimates for $\theta$ are then\footnote{Obviously this is equivalent to $\inprod{\mathbf{W}_{d_g \times d_g}^{1/2} \mathbf{g}_l(\theta)}{\mathbf{W}_{d_g \times d_g}^{1/2} \mathbf{g}_l(\theta)} = \mathbf{g}_l(\theta)^\top \mathbf{W}_{d_g \times d_g} \mathbf{g}_l(\theta)$, \glssymbol{transpose}[$\mathbf{g}^\top$] being the transpose.}
\begin{equation}\label{eq: GMM est}
\theta_{GMM}^* =\underset{\theta}{\mathrm{argmin}}{ \lVert \mathbf{W}_{d_g \times d_g}^{1/2} \mathbf{g}_l(\theta) \rVert } .
\end{equation}
The most statistically efficient matrix is $\mathbf{W}_{d_g \times d_g}=\expect[ \mathbf{g}_l(\theta_o) \circ \mathbf{g}_l(\theta_o) ]^{-1} $, which is the inverse of the covariance matrix of our moment conditions. We can prove that such estimators exhibit asymptotic normality such that $ \sqrt{l} (\theta_{GMM}^* - \theta_o) \sim \mathcal{N}(0, (\mathbf{G_{d_g \times d_{\theta}}^\top W_{d_g \times d_g} G_{d_g \times d_{\theta}}})^{-1} )$, where $\theta_o$ is the true parameter value and 
\begin{equation}
\mathbf{G}_{d_g \times d_{\theta}}=\expect\left[\frac{\partial \mathbf{g}(\theta) }{ \partial \theta } \Big\vert_{\theta=\theta_o} \right] 
\end{equation}

\noindent Where the expectation is w.r.t. $\mathbf{X}$ and the derivative $\frac{d}{d\theta}$ is understood as the gradient in the case $d_{\theta}>1$. Interestingly, this suggests that the correct moment conditions can recover the same information that was ``thrown away'' from the MLE. Indeed, we will see that the result of \cite{carrasco2000generalization} formalizes just this idea to construct CGMM estimators. For now, let us consider the moment condition of form  
\begin{equation}\label{eq: score}
\mathbf{g}(\theta : \mathbf{x}_r) = \dfrac{\partial \ln( f_{\mathbf{X}}(\mathbf{x}_r|\theta ))}{\partial \theta} .
\end{equation}
That is, Eq. (\ref{eq: score}) is the score function. Obviously, $\hat{\theta}_{MLE}$ will solve Eq.\ (\ref{ind moment cond}), while $\mathbf{G}_{d_{\theta} \times d_{\theta}}= -\mathbf{I}(\theta)$ and $\mathbf{W}_{d_{\theta} \times d_{\theta}}=\mathbf{I}(\theta)$ where $\mathbf{I}(\theta)$ is the Fisher information matrix.
Therefore $ \sqrt{l} (\theta_{GMM}^* - \theta_o) \sim \mathcal{N}(0, \mathbf{I}^{-1} )$. That is, the GMM estimators achieve the \gls{cr} bound and are thereby as efficient as the MLEs. The intuition here is unsurprising as we are making use of the same information, namely specification, as in the MLE. In some sense, then, the GMM can be seen as more general than likelihood estimation. The CGMM in turn is a further generalization to a \textit{continuum} of moment conditions. 

\subsection{The CGMM}

Consider the specific problem of estimating models efficiently from their integral transforms, namely the aforementioned Tweedie and Stable models. In \cite{feuerverger1981some}, the authors suggested the following moment condition. For $\lbrace \bm{\tau}_k \rbrace_{k=1}^{d_g}$ we choose a moment condition involving a CF of the form
\begin{equation}\label{eq: mom cond}
g_{k}(\theta : \mathbf{x}) =  e^{i \inprod{\bm{\tau}_k}{ \mathbf{x} } } - \phi_{\mathbf{x}|\theta}(\bm{\tau}_k).
\end{equation}
They claimed this could be thought of as a potentially infinitely over-specified moment condition. The idea is that by sampling an increasing number of $\bm{\tau}$ values and proceeding with the typical GMM, one could achieve the CR bound. However, they did not prove this result, nor did they take into account how to construct the proper covariance matrix $\mathbf{W}$ in such a way as to continuously match the infinite-dimensional moment condition of Eq.\ (\ref{eq: mom cond}). 

To see why a continuous moment matching is necessary consider the resulting vector-valued function $\mathbf{g}_l(\theta)$ (Eq.\ (\ref{eq: pop moment cond}) and covariance matrix: 

\begin{align*}
\mathbf{K}_{d_g \times d_g}=\mathbf{W}_{d_g \times d_g}^{-1} =\expect[ \mathbf{g}_l(\theta_o) \mathbf{g}_l(\theta_o)^\top ] = { \left\lbrace k_{ij} \right\rbrace }_{i,j=1}^l 
\end{align*}
where 
\begin{align*}
k_{ij} = \phi_{X|\theta_o}( \tau_i - \tau_j ) - \phi_{X|\theta_o}( \tau_i )\phi_{X|\theta_o}( - \tau_j ) .
\end{align*}
Note that $\mathbf{K}_{d_g \times d_g}$ is Hermitian and positive definite, and satisfies properties that we would normally find desirable. However, one can show that the smallest eigenvalue of such a matrix will approach zero as $d_g$ goes to infinity. This will make operations involving $\mathbf{W}_{d_g \times d_g}$ and the GMM very unreliable numerically. In fact, if we were to take the limit as $d_g$ approaches infinity, $\mathbf{W}_{d_g \times d_g}$ would become unbounded and not invertible. This leaves us in need of a procedure to \textit{continuously} match our $\mathbf{W}$-norm in (\ref{eq: GMM est}) with the moment conditions. 

In \cite{carrasco2000generalization} the authors tackled these problems by inventing the CGMM. Rather than simply sampling from a continuous object as in  Eq.\ (\ref{eq: mom cond}), they reformulated the GMM in \textit{continuous} terms: vectors and matrices become elements of a Hilbert space and operators, respectively. Intuitively, if we consider $d_g \longrightarrow \infty$ instead of a \textit{vector} moment condition we get a \textit{function}, so that $\mathbf{g}_l(\theta) \longrightarrow \frac{1}{l} \sum_{r=1}^l e^{i \inprod{\bm{\tau}}{ \mathbf{x}_r} } - \phi_{X|\theta}(\bm{\tau}) $, specifically a function of $\bm{\tau}$. 

In the interest of brevity, given a function of several variables $g$ we will write integrals like so:
\begin{equation}
\int g(\mathbf{x}) d\mathbf{x} = \underset{R^n}{\int ...\int} g(x_1,..x_n) dx_1...dx_n .
\end{equation}
\noindent To avoid confusion with the discrete inner product above, the $L^2$ inner product for functions $f$ and $g$ is defined as
\begin{equation}
< g , f >_{L^2} =  \int g(x) \overline{f(x)} dx.
\end{equation}
\noindent In order to give the typical notion of $L^2$, we set
\begin{equation}
L^2 = \left\lbrace f \Bigg| < f, f >_{L^2}  < \infty \right\rbrace,
\end{equation}
the space over which we will primarily be working in the CGMM. Finally, given $f\in L^2$ and a kernel function $k: \mathbb{R}^{d_{\theta}} \times \mathbb{R}^{d_{\theta}} \rightarrow  \mathbb{R}$ we will denote the notion of a \glssymbol{f_op}[Fredholm-style operator $\mathcal{K}$] as
\begin{equation}
(\mathcal{K} f)(\mathbf{t}) = \int k( \mathbf{t},\mathbf{s} ) f(\mathbf{s}) d\mathbf{s}.
\end{equation}

By way of example to illustrate the CGMM, we again consider an i.i.d.\ sample $\lbrace x_1, ..., x_l \rbrace$, but this time define the moment condition by a \textit{function} of $\tau$:  
\begin{equation}
h(\bm{\tau},\theta : \mathbf{x}) = e^{i \inprod{\bm{\tau}}{\mathbf{x} }} - \phi_{\mathbf{x}|\theta}(\bm{\tau})  
\end{equation}
so that 
\begin{equation}\label{eq: CF moment cond}
h_l(\bm{\tau}, \theta) = \frac{1}{l} \sum_{r=1}^l h(\bm{\tau},\theta : \mathbf{x}_r).
\end{equation}
And rather than a matrix we define a positive-definite kernel function:
\begin{equation}\label{eq: kernel}
k(\bm{\tau}_1,\bm{\tau}_2) = \expect[ h_l(\bm{\tau}_1, \theta) \overline{h_l(\bm{\tau}_2, \theta)} ] .
\end{equation}
We can generalize the GMM estimators to their continuous counterparts: 
\begin{equation}\label{eq: CGMM Obj}
\theta_{CGMM}^* = \underset{\theta}{\mathrm{argmin}}{ \lVert \mathcal{K}^{-1/2} h_l \rVert_{L^2} } = \underset{\theta}{\mathrm{argmin}}{ < \mathcal{K}^{-1/2} h_l, \mathcal{K}^{-1/2} h_l >_{L^2} } .
\end{equation}
While the operator $\mathcal{K}$ and its inverse are self-adjoint, the expression $< h_l , \mathcal{K}^{-1} h_l >_{L^2} $ is often undefined. The inverse square root of the operator has a larger domain than the simple inverse and that is why we write Eq.\ (\ref{eq: CGMM Obj}) as here. 

Estimators of the kind defined in Eq.\ (\ref{eq: CGMM Obj}) can be shown to be \textit{as efficient} as the MLE, achieving the CR bound. The intuition behind this is discussed in \cite{carrasco2014asymptotic}. The GMM can recover the MLE if the moment conditions involve the score; by generalizing the GMM to Hilbert spaces we can relax this. We now only require the new moment conditions to contain the score within their linear \textit{closure}. Naturally, this is true for moment conditions of form (\ref{eq: CF moment cond}). Figure \ref{fig: comparison} summarizes the situation.

\begin{minipage}{0.45\textwidth}
\begingroup
\addtolength{\jot}{1em} 
\begin{align*}
\mathbf{g}(\theta: \mathbf{x}) &= (g_1(\theta :\mathbf{x}),...,g_{d_g}(\theta :\mathbf{x})) \\
\mathbf{g}_l(\theta) &= \frac{1}{l} \sum_{r=1}^l \mathbf{g}(\theta: \mathbf{x}_r) \\
\mathbf{g}_l(\theta)&:\hspace{2mm}
 \begin{matrix}
\mathbb{R}^{d_{\theta}} & \longrightarrow & \mathbb{R}^{d_g} \\
\theta & \longrightarrow & \mathbf{g}_l(\theta) \\
\end{matrix}\\
\mathbf{W}_{d_g \times d_g} &:\hspace{2mm}
 \begin{matrix}
\mathbb{R}^{d_g} & \longrightarrow & \mathbb{R}^{d_g} \\
\mathbf{g}_l & \longrightarrow & \mathbf{W}\mathbf{g}_l \\
\end{matrix} \\
\theta_{GMM}^* &= \underset{\theta}{\mathrm{argmin}}{ \lVert \mathbf{W}_{d_g \times d_g}^{1/2} \mathbf{g}_l(\theta) \rVert }
\end{align*}
\endgroup
\end{minipage}%
\hfill\vline\hfill
\begin{minipage}{0.45\textwidth}
\begingroup
\addtolength{\jot}{1em} 
\begin{align*}
h(\tau,\theta : \mathbf{x}) &= e^{i \inprod{ \bm{\tau} }{ \mathbf{x} } } - \phi_{\mathbf{x}|\theta}(\bm{\tau}) \\
 h_l(\bm{\tau}, \theta) &= \frac{1}{l} \sum_{r=1}^l h(\bm{\tau},\theta : \mathbf{x}_r) \\
 h_l(\bm{\tau}, \theta)&:\hspace{2mm}
 \begin{matrix}
\mathbb{R}^{d_{\theta}} & \longrightarrow & L^2 \\
\theta & \longrightarrow & h(\bm{\tau},\theta)  \\
\end{matrix} \\
\mathcal{K} &:\hspace{2mm}
 \begin{matrix}
L^2 & \longrightarrow & L^2 \\
h(\bm{\tau},\theta) & \longrightarrow & \int k( \mathbf{s},\bm{\tau} ) h(\bm{\tau},\theta) d\bm{\tau} \\
\end{matrix} \\
\theta_{CGMM}^* &= \underset{\theta}{\mathrm{argmin}}{ \lVert \mathcal{K}^{1/2} h_l \rVert_{L^2} }
\end{align*}
\endgroup
\end{minipage}
\begingroup
\captionof{figure}{Summary of GMM and CGMM counterparts}\label{fig: comparison}
\endgroup 
\vspace*{5 mm}

Evaluating the objective is equivalent to solving a Fredholm integral equation of the first kind. That is, the function $u(\tau)=(\mathcal{K}^{-1/2}h_l)(\tau)$ can be seen as a solution to the equation 
\begin{equation}\label{eq:fred eq}
\mathcal{K}^{1/2} u = h_l ,
\end{equation}
where $\mathcal{K}: D(\mathcal{K}) \rightarrow R(\mathcal{K})$ is the integral operator with kernel given by Eq.\ (\ref{eq: kernel}).

In what follows, we provide a quick introduction to solving integral equations in a practical way. In general, the true inverse of an operator, namely $\mathcal{K}^{-1}: R(\mathcal{K}) \rightarrow D(\mathcal{K})$ such that $(\mathcal{K}\mathcal{K}^{-1} f)(\cdot)=id(\cdot)$, is unbounded. Solutions of Eq.\ (\ref{eq:fred eq}) will only be defined for a dense subset of $D(\mathcal{K})$. To ensure existence everywhere, we need to weaken our notion of a solution to one that solves the least squares problem. Therefore, rather than solving $\mathcal{K}u=f$, we solve 
\begin{equation}\label{eq:PI}
\mathcal{K}^{\dagger}u=\operatornamewithlimits{argmin}_{u \in D(\mathcal{K})} \left \lbrace \lVert \mathcal{K}u - f \rVert^2 \right\rbrace .
\end{equation}
The operator $\mathcal{K}^{\dagger}:R(K) \rightarrow D(K)$ is called the \textit{pseudoinverse} of $K$. This satisfies the existence problem. Unfortunately, problems of the kind (\ref{eq:PI}) are still not well posed. Given especially that our moment conditions (\ref{eq: CF moment cond}) are estimated from data, we need to introduce the regularized problem that \textit{is} well posed: 
\begin{equation}\label{eq:regularization}
\mathcal{K}^{\dagger}u_{\lambda}=\operatornamewithlimits{argmin}_{u \in D(\mathcal{K})} \left \lbrace \lVert \mathcal{K}u - f \rVert^2 + \lambda\lVert u \rVert^2 \right\rbrace .
\end{equation}
This regularized minimization problem has a unique solution $u_{\lambda}=(\mathcal{K}'\mathcal{K}+\lambda I)^{-1} \mathcal{K}' f$, where $\mathcal{K}'$ is the adjoint of $\mathcal{K}$. As $\lambda \rightarrow 0$, we recover (\ref{eq:PI}). We can thus think of the regularized problem as approximating an ill-posed problem with a ``nearby" well-posed one. 

\subsection{Loss Reserve Estimates}\label{LR est}

Let us revisit the example from Section \ref{sec: LossRes} involving the Gamma losses in Figure \ref{fig: Tweedie Ex}. We will compute estimates for $\eta$, $\nu$ and $\gamma$ using the CGMM instead of the MLE. We define the following: 
\begin{itemize}

\item $X_{i,j} \sim \text{Gamma} \left( \frac{1}{\gamma_j} , \eta_i \nu_j \gamma_j \right)$ 

\item $ \theta =(\eta_i, \nu_j, \gamma_j )$ so that  $\phi_{X_{i,j} |\theta }(\tau) = \expect[e^{ i\tau X_{i,j} }] =  ( 1- i\tau \eta_i \nu_j \gamma_j )^{ - \frac{1}{\gamma_j} }$

\item $h_{i,j}(\tau) =  e^{ i \tau X_{i,j}} - \phi_{X_{i,j} |\theta }(\tau)$

\item $\mathcal{K}_{i,j} f = \int k_{i,j}(\tau,s) f(s) d\pi(s)$ where  $k_{i,j}(\tau,s)= \phi_{X_{i,j} |\theta }( \tau - s ) - \phi_{X_{i,j} |\theta }( \tau )\phi_{X_{i,j} |\theta }( - s )$
\end{itemize}

\noindent The objective function to be used is 

\begin{equation}\label{eq: LR_CGMM_obj}
(\hat{\eta_i}, \hat{\nu_j},\hat{\gamma}_{i,j}) = \operatornamewithlimits{argmin}_{\eta_i, \nu_j, \gamma_j}  \sum_{i,j}  < \mathcal{K}_{i,j}^{-1/2}\hat{h}_{i,j} , \mathcal{K}_{i,j}^{-1/2}\hat{h}_{i,j}  > . 
\end{equation}

Evaluating the $\mathcal{K}_{i,j}^{-1/2}\hat{h}_{i,j}$ using the Nyström method and numerically optimizing (detailed in the next section), we obtain the mean expected claims shown in Table \ref{tab:CGMM gamma results}. 

\begin{table}[ht]
\centering
\begin{tabular}{|c|c|c|c|}
\hline
Accident Year & Realized Losses & Estimated Claims (CGMM) & Estimated Claims (MLE) \\ \hline
2             & 2.92            & 1.93                                & 2.66                               \\ \hline
3             & 8.10            & 9.70                                & 7.88                               \\ \hline
4             & 12.37           & 20.08                               & 17.04                              \\ \hline
5             & 27.07           & 37.22                               & 27.92                              \\ \hline
6             & 41.73           & 71.03                               & 45.49                              \\ \hline
7             & 63.84           & 85.42                               & 66.37                              \\ \hline
8             & 94.94           & 100.68                              & 98.76                              \\ \hline
9             & 127.07          & 110.27                              & 113.36                             \\ \hline
10            & 167.27          & 134.52                              & 138.03                             \\ \hline
\end{tabular}
\caption{Realized losses from the scenario used in Figure \ref{fig: Tweedie Ex} vs the MLE- and CGMM-estimated mean losses }
\label{tab:CGMM gamma results}
\end{table}

\begin{figure}[ht]
  \centering
  \includegraphics[scale=0.6]{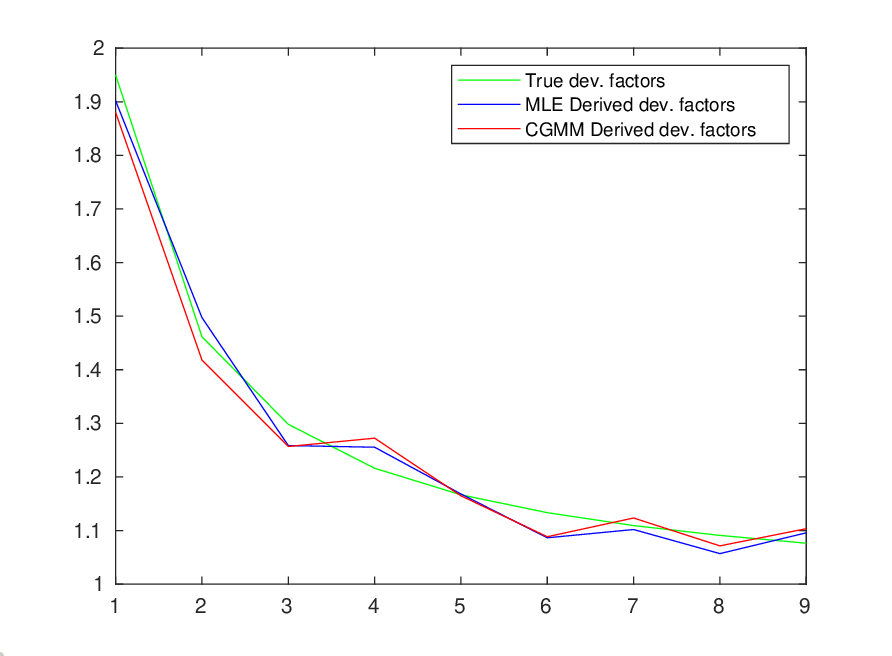}
  \caption{Comparison of true estimated development factors from the Gamma model (green) vs MLE model (blue) and CGMM estimates from Eq.\ \eqref{eq: LR_CGMM_obj} (red)}
 \label{fig: CL plot}
\end{figure}

\noindent Given that for Tweedie losses with power parameter close to $2$, we expect the CL and MLE estimates to be close, it is unsurprising that the CGMM parametric estimate for the development factors is also similar, as shown in  Figure \ref{fig: CL plot}.

\subsection{Numerical Considerations}\label{sec: numerical CGMM}

\paragraph{Estimating the Kernel Function}

In both the GMM and the CGMM, we have thus far ignored the estimation of the matrix $\mathbf{W}_{d \times d}$ and the kernel operator $\mathcal{K}$ respectively. Recall they are defined as the expectation of the Kronecker product of our moment functions, which require the very quantity we are trying to estimate, namely $\theta_o$. In the i.i.d.\ case, the natural choice of estimator is simply
\begin{equation}\label{iid kernel}
\hat{k}(\tau_1,\tau_2) = \frac{1}{l} \sum_{t=1}^{l}(e^{i\tau_1 x_t} - \hat{\phi}(\tau_1))\overline{(e^{i\tau_2 x_t} - \hat{\phi}(\tau_2))},
\end{equation}
where $\hat{\phi}(\tau)=\frac{1}{l} \sum_{t=1}^{l}e^{i\tau x_t}$.

Unfortunately, in loss reserving we do not have the luxury of sizable i.i.d.\ samples. To address this issue, we continuously update the kernel function (defined in (\ref{eq: kernel})) after a reasonable initial estimate (e.g.\ the CL estimates). This strategy is appropriate for the same reason that continuously updating in the discrete GMM case works: as \cite{carrasco2000generalization} shows, any appropriate kernel function that produces a norm can produce a consistent estimate for $\theta_o$. Unfortunately, this does add some computational cost. 

\paragraph{Estimation of $\lambda$} The choice of regularization parameter $\lambda$  obviously affects the estimators. If it is too large (resp.\ too small), the problem we are solving is too far from the original regularized problem (resp.\ we lose numerical stability). We direct the reader to the discussion in \cite{kotchoni2012applications} for possible simulation-based approaches to optimizing this parameter. In the loss-reserving setting, we find that a good ``rule of thumb'' is to set $\lambda \approx 10^{-7}$ when the number of development years is around $10$. 

\paragraph{Numerical Integral Equation Solutions 1: The Nyström method} The most obvious way of evaluating (\ref{eq:fred eq}) is via numerical quadrature. We introduce quadrature points\footnote{Capital $Q$ here is taken to be deterministic. } $\lbrace s_q \rbrace_{q=1}^Q$ and corresponding weights $\lbrace w_q \rbrace_{q=1}^Q$ such that
\begin{align*}
(\mathcal{K} h)(\tau) =\int k(\tau,s) h(s) d\pi(s) \approx \sum_{q=1}^Q w_q k(\tau,s_q) h(s_q) .
\end{align*}
We can then represent the continuous problem of the kind (\ref{eq:fred eq}) as  
\[
\begin{pmatrix}
w_1 k(s_1,s_1) & \dots & w_q k(s_1,s_q)  & \dots & w_Q k(s_1,s_Q) \\
\vdots & & \vdots & & \vdots\\
w_1 k(s_q,s_1) & \dots & w_q k(s_q,s_q)  & \dots & w_Q k(s_q,s_Q) \\
\vdots & & \vdots & & \vdots\\
w_1 k(s_Q,s_1) & \dots & w_q k(s_Q,s_q)  & \dots & w_Q k(s_Q,s_Q) 
\end{pmatrix}^{1/2}
\begin{pmatrix}
u(s_1) \\
\vdots \\
u(s_q) \\
\vdots \\
u(s_Q) \\
\end{pmatrix}
=
\begin{pmatrix}
\hat{h}_T(s_1) \\
\vdots \\
\hat{h}_T(s_q) \\
\vdots \\
\hat{h}_T(s_Q) \\
\end{pmatrix} 
\]
or more succinctly as $\mathbf{K}_{Q \times Q}^{1/2} \mathbf{u}_{Q \times 1} = \mathbf{h}_{Q \times 1}$. We then evaluate the CGMM objective (\ref{eq: CGMM Obj}) as $\inprod{\mathbf{u}_\lambda}{\mathbf{u}_\lambda}$, where $\mathbf{u}_\lambda = (\mathbf{K}_{Q \times Q}+\lambda \mathbf{I}_{Q \times Q} )^{-1} \mathbf{K}_{Q \times Q}^{1/2} \mathbf{h}_{Q \times 1} $.

\paragraph{Numerical Optimization}

Consider a more general moment function of the form
\begin{equation}\label{general moment fun} 
h(\tau,\theta : X) = m(X,\tau) - E[m(X,\tau)].
\end{equation}
Thus far, we have only discussed expressions of the kind $m(X,\tau)=e^{ i \tau X }$ leading to $E[m(X,\tau)]=\phi_{X|\theta}(\tau)$, the CF. The CGMM is not limited to such expressions. For example, given a CDF $F_X (x)$ it is just as natural to choose $m(X,\tau)=\mathbf{1}_{ \lbrace X>\tau \rbrace }$, leading to an equivalent expression of form
\begin{equation}\label{cdf moment fun} 
h(\tau,\theta : X) = \mathbf{1}_{ \lbrace X>\tau \rbrace } - F_X (\tau) .
\end{equation}

A closed-form CDF is, of course, lacking in the models we are interested in, so this point may seem moot. However, we emphasize that if the linear closure of expressions of the form $h(\tau,\theta : X)$ contains the score of our distribution, we can employ the CGMM confidently, as demonstrated by the following theorem.

\begin{theorem}[\cite{carrasco2014asymptotic}]\label{thm:closure}
Consider the subspace $L^2(h)$ of $L^2$ formed by the linear hull of $\{ h(\tau,\theta : X), \tau \in \mathbb{R}^{d_x} \}$. That is, for $w_j \in \mathbb{R}$ consider the space of random variables of the form
\begin{align*}
j_n=\sum_{j=1}^n w_j  h(\tau_j,\theta : X),
\end{align*}
and their mean square limits $G$ such that
\begin{align*}
\expect[\Vert j_n - j \Vert_{L^2}] \xrightarrow[n \to \infty]{} 0 .
\end{align*}
If the score $s_{\theta}(x) = \dfrac{\partial \ln( f_{X}(x|\theta ))}{\partial \theta} \in L^2(h)$, then the resulting CGMM estimator for $\theta$ is asymptotically as efficient as the MLE. 
\end{theorem}

Thus, for example in one dimension we can use the characteristic function and approximate a limit point in $L^2(h)$ by an integral as follows: 
\begin{align*}
\lim_{n\to\infty} \sum_{j=1}^n w_j  h(\tau_j,\theta : x) & \approx \int h(\tau,\theta : X) w(\tau) d\tau \\
&= \int e^{i\tau x} w(\tau) d\tau - \int \phi_{X|\theta}(\tau) w(\tau) d\tau \\
& \qquad\qquad \left(\text{ letting $w(\tau)=\frac{1}{2\pi}\int e^{-i\tau x} s_{\theta}(x) dx $ }\right) \\ 
&= s_{\theta}(x) - \int s_{\theta}(x) \left[ \frac{1}{2\pi}\int e^{-i\tau x} \phi_{X|\theta}(\tau) \right] dx \\
&=s_{\theta}(x) - E[ s_{\theta}(x) ] \\
& \qquad\qquad  \text{ (where the expectation is w.r.t.\ $\theta$, i.e.\ $E[ s_{\theta}(x) ]=0$) }\\ 
&= s_{\theta}(x).
\end{align*}

\begin{figure}[h!]
\hspace*{-1.5cm}
  \centering
  \subfloat[]
  {
  \includegraphics[scale=0.55]{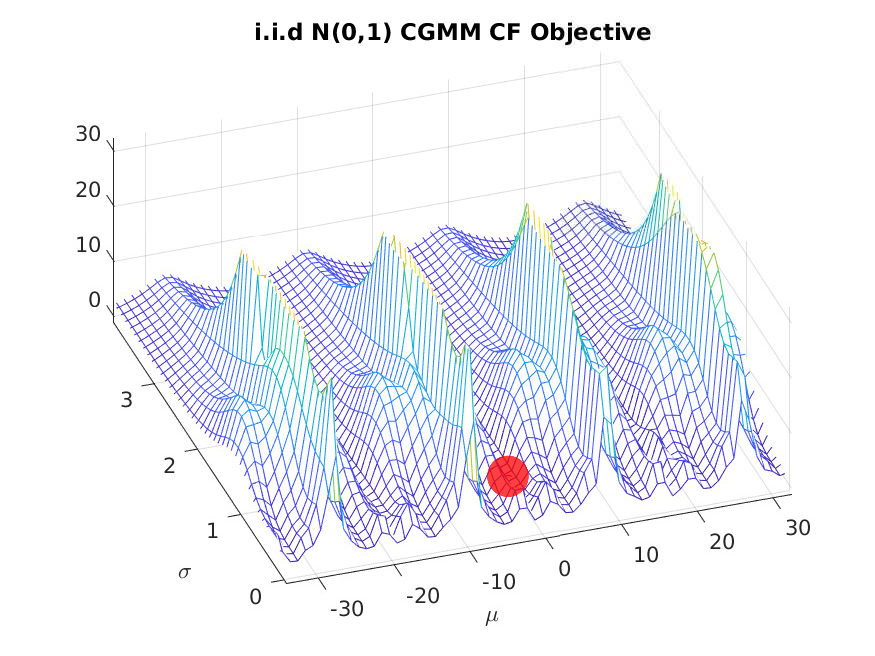}\label{fig: 3D fig}
  }
  \subfloat[]
  {
  \includegraphics[scale=0.55]{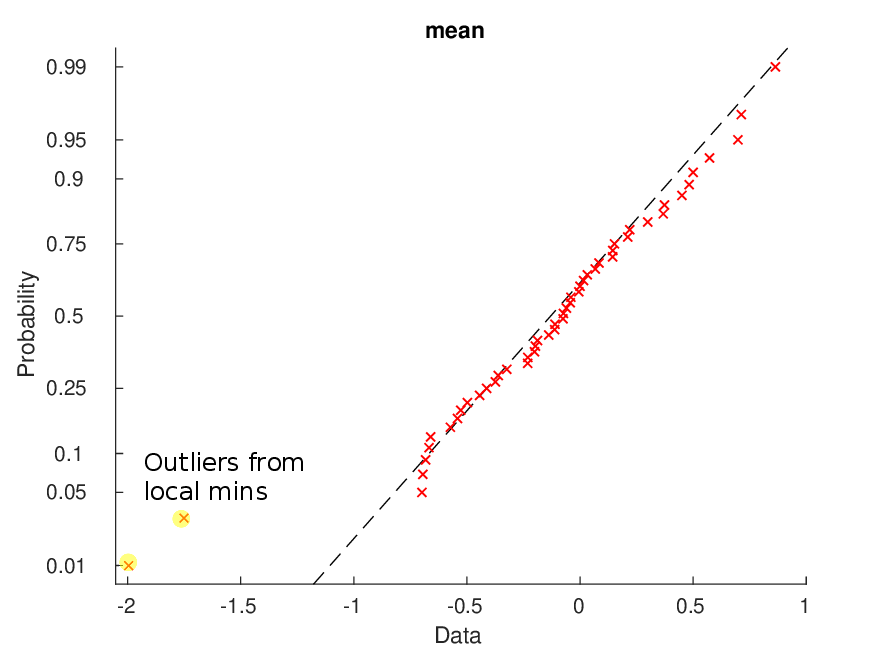}\label{fig: outlier PP plot}
  }
  \caption{Complex-valued functions lead to periodicity and outliers}\label{fig: Obj vs Outlier}
\end{figure}

Theoretically, there are few issues with choosing the CF. Practically speaking, however, the CF may lead to non-convex objectives that are difficult to optimize. For instance, consider the following simple situation. Given an i.i.d.\ sample of $\lbrace x_1, ..., x_l \rbrace$, where $X \sim \mathcal{N}(\mu,\sigma^2)$ with $\mu=0$ and $\sigma=1$, and $l=10$ and $m(x,\tau)=e^{ i\tau x }$, we can plot the surface of (\ref{eq: CGMM Obj}) as shown in Figure \ref{fig: 3D fig}. Keep in mind, we wish to find the minimum.

The true minimum (shown by the red dot in Figure \ref{fig: 3D fig}) is still close to $(0,1)$, as expected. We can see however that the presence of complex-valued functions creates oscillations and periodicity. Often, numerical errors and the sheer non-convexity of the objective can lead to standard gradient descent algorithms getting ``stuck" out in neighbourhoods of adjacent local minimums. This is especially true in the heavier-tailed case, where it is harder to guess the initial point to start the optimization. To see this, look for instance at the Prob-Prob plot of the sampling distribution for $\mu$ (generated using ($l=10$)-sized Stable samples with $\alpha=1.5$) in Figure \ref{fig: outlier PP plot}. While the distribution appears mostly Normal (which, in theory, would be the case asymptotically) we can see that these numerical issues create a few notable outliers. In  \cite{kotchoni2012applications} the application of the CGMM to heavy-tailed Stable variates required rather large sample sizes to overcome this problem. We suspect that this issue is partly why the CGMM has not yet seen wide application. 

In the loss-reserving case, the picture is more challenging. The increased dimensions of the objective and possible local minimums mean that it is a near certainty that standard algorithms will get stuck. Two similar alternative approaches to finding a global minimum that we explore are MATLAB's built-in scatter-search and genetic algorithms. The former is used to generate the results in our Gamma example of Section \ref{LR est}. Both of these algorithms search a larger portion of the feasible space and often produce favourable results. Another useful trick is to add a penalty for solutions found too far away from a good first estimate (say, the chain-ladder parameters). These approaches often produce the true optimum. However, the reality is that the fundamental non-convexity and periodicity of the objective are too much to overcome reliably. 

Ideally, we would like to find moment conditions that are not just statistically valid but also produce well-behaved objective functions. It is preferable to make use of integral transforms that produce real and (log-)convex functions such as the moment-generating function. That is,
\begin{equation}\label{eq: MGF moment function}
h_t(\tau,\theta:X)=e^{\tau X} - M_{X|\theta}(\tau) 
\end{equation}
with corresponding kernel function 
\begin{equation}\label{eq: MGF kernel}
k(s,\tau) = M_{X|\theta}(s+\tau)-M_{X|\theta}(s)M_{X|\theta}(\tau) 
\end{equation}
where $M_X(\tau)$ is the \gls{mgf_words} of $X$. However, we run into issues showing this satisfies the conditions of Theorem \ref{thm:closure}. To begin with , if left unrestricted, $h$ is no longer in $L^2$. Additionally, we would require $w(\tau)=-\frac{i}{2\pi} \int e^{-tx} s(x) dx$, which has no real coefficients. 

In light of this, we can instead perform a ``Wick rotation" and recover the MGF that way. Setting $\tau \rightarrow -i\tau$, we can make use of (\ref{eq: MGF moment function}) as long as we keep the CF form for $\mathcal{K}$:
\begin{equation}\label{eq: Wick kernel}
k(s,\tau) = M_{X|\theta}(s-\tau)-M_{X|\theta}(s)M_{X|\theta}(-\tau) .
\end{equation}

\begin{figure}[h!]
  \centering
  \includegraphics[scale=0.8]{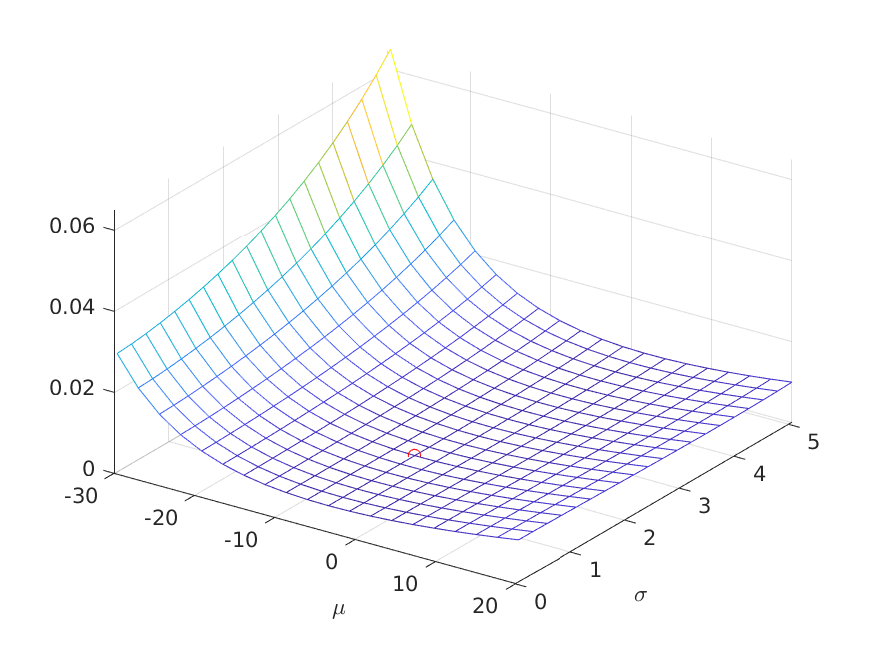}
  \caption{CGMM objective using the MGF (\ref{eq: MGF moment function}) and the kernel from (\ref{eq: Wick kernel})}
 \label{fig: 3D fig MGF}
\end{figure}

Using the MGF and this new kernel, we can generate a new objective function as in Figure \ref{fig: 3D fig MGF} from the same scenario and data used in Figure \ref{fig: 3D fig}. As we can see, the objective function is now convex and much more amenable to numerical optimization. 

One aspect of note that will reoccur in Section \ref{sec: muti lob} is that the objective function has little curvature around the optimum. This ``flatness'' becomes even more pronounced as we add dimensions to the data (or lines of business in the loss-reserving problem). 

Before continuing to numerical examples, we have one additional issue to consider. To guarantee our conditions are properly bounded, we introduce a dampening function. In methods where the empirical characteristic function is used (e.g.\ kernel density estimation), it is well-known that some values of $\tau$ are unreliable. If we use the MGF instead, the same issue arises but now the actual MGF may not even be defined at some $\tau$. As a remedy to both issues, dampening functions are employed to emphasize more reliable regions of $\tau$ over others. In our setting, this takes the form
\begin{equation}\label{eq: damped general moment fun} 
h_t(\tau,\theta:X) = \Psi(\tau) (m(X,\tau) - E[m(X,\tau)]) 
\end{equation}
where $\Psi(\tau)$ is a dampening function. It is easy to show that this fits well into the CGMM framework.  For instance, in \cite{carrasco2000generalization} this takes the form of a new measure $\pi(\mathbf{x})$ and space $L^2(\pi) = \lbrace f | f(\mathbf{x})\overline{f(\mathbf{x})} d\pi(\mathbf{x}) < \infty \rbrace$ where the CGMM results still hold true. 


\section{Simulation Results}\label{sec: LossSims}

\subsection{Single LoB Illustration}\label{sec: 5.1}

For each of the examples discussed in this section, we simulate 50 single LoB (line of business) loss triangles with 10 accident years and 10 development years. For simplicity, we use the same parameter values as in our Gamma examples. We use (\ref{eq: MGF moment function}) and (\ref{eq: Wick kernel}) to construct a CGMM objective of the form (\ref{eq: LR_CGMM_obj}). We then find the estimators via MATLAB's \texttt{fmincon} function, which employs sequential quadratic programming.

\paragraph{Tweedie Losses: Compound Gamma and Gamma} In the univariate case, the (reproductive) MGF of the Tweedie distribution is
\begin{equation}
M_{X|\mu,\sigma}(\tau)=\exp\left( \frac{1}{\sigma^2} [\kappa_p (g(\mu) +\tau \sigma^2 )-\kappa_p (g(\mu))] \right) .
\end{equation}

For some values of $p$, the MGF cannot be defined past some positive $\tau$. For example, in the $p=2$ (Gamma) case, we require $\tau<\frac{1}{\mu \sigma^2}$. This may create some issues with the kernel function (recall the $s-\tau$ argument in the first term). To avoid this we choose a dampening function of the form $\Psi(\tau) = \mathbf{1}_{\{ t<0 \}}$ and consider the MGF only over $(-\infty,0]$:
\begin{equation}\label{eq: tweedie h}
h(\tau,\theta: x) = \mathbf{1}_{\{ \tau<0 \}}\left( \exp\{ \tau x \} - M_{X|\theta}(\tau) \right) .
\end{equation}

Using this formulation, we construct an objective of the form (\ref{eq: LR_CGMM_obj}), but rather than the CF we use the MGF and a kernel of the form (\ref{eq: MGF kernel}). Integration for $\lVert \mathcal{K}^{-1/2} \hat{h}_T \rVert$ is done via a simple trapezoidal scheme \footnote{Interestingly, Newton--Coates schemes require fewer quadrature points than Gaussian quadrature. Despite having the freedom to select any $\pi$, it appears the function is well-behaved enough that linear approximations are sufficiently precise. }

Table \ref{tab:Tweedie 1 LoB table} summarizes the statistics for the CGMM estimates for Tweedie power parameters $p=1.2$ and $p=2$. For comparison, we include the chain ladder and MLE estimators, respectively. While the mean parameters of $\eta$ and $\nu$ are slightly biased, the advantage goes to the CGMM in the estimation of the scale. The bias is present in the Stable example as well and we will discuss why in that case. For now, it is likely safe to conclude that the CGMM and CL/MLE estimators are at least comparable.  

\paragraph{Stable Losses} Given that Stable variables do not admit even second moments for all but $\alpha<2$, using the MGF over the CF to generate a CGMM objective in the Stable case seems misguided at first glance. Fortunately, \cite{eaton1971extreme} and \cite{samoradnitsky_book} have studied the Laplace transformation of an extreme Stable distribution ($\beta=-1$) and applied the Paley–Wiener theorems to conclude that for $X\sim S_{\alpha}(\mu,\sigma,-1,)$ the ``moment generating function" is given by
\begin{equation}
M_{X|,\mu,\sigma}(\tau)=\exp\left( \mu\tau - \frac{\sigma^\alpha \tau^\alpha}{\cos(\alpha \pi / 2)} \right) , \text{ for } \tau>0 .
\end{equation}

\noindent Note the fact that the resulting cumulant function is similar to the Tweedie, which formalizes the connection we hinted at earlier. 

Again, for negative values of $\tau$ this obviously produces complex-valued arguments. Hence, once again we consider only the half-line that works for us. We also multiply by $-1$ to ensure positive losses:
\begin{equation}\label{eq: tweedie h}
h(\tau,\theta: x) = \mathbf{1}_{\{ \tau<0 \}}\left( \exp\{ \tau (-x) \} - M_{-X|\theta}(\tau) \right) .
\end{equation}

\begin{figure}[h!]
  \centering
  \subfloat[MGF-based Objective]
  {
  \includegraphics[scale=0.5]{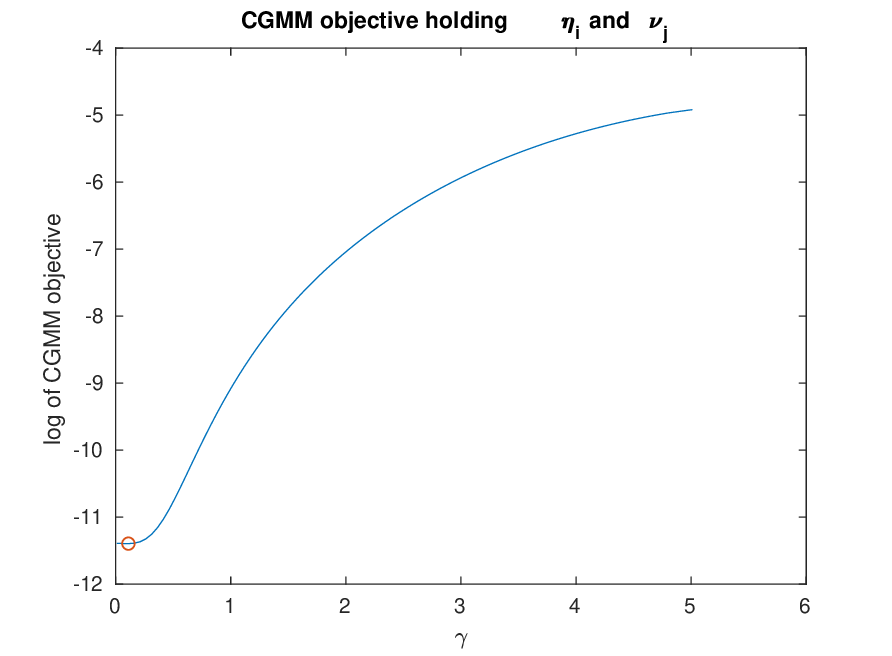}\label{fig:f1}
  }
  \hfill
  \subfloat[CF-based Objective]
  {
  \includegraphics[scale=0.5]{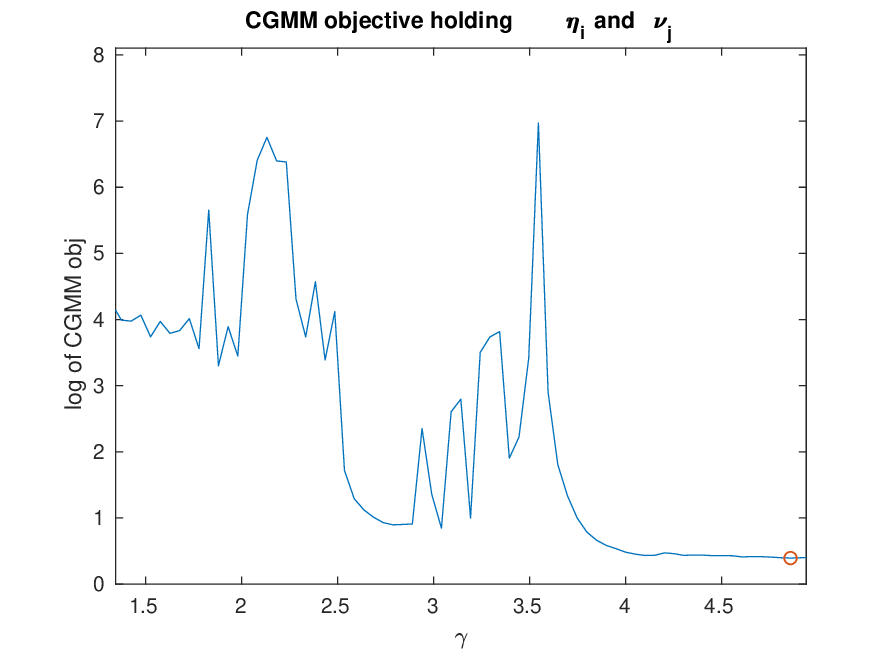}\label{fig:f2}
  }
  \caption{Samples of the CGMM objective functions around the minima for MGF and CF moment functions}
  \label{fig:MGFvsCF}
\end{figure}

In Figure \ref{fig:MGFvsCF}, we compare a slice of the CGMM objective using the Stable MGF with another using the CF. As one can see, the difference is quite stark. Again, while a global search may give the correct minimum, it is not difficult to see why the MGF-based moment functions give much better and more consistent performance.

\begin{figure}
  \centering
  \includegraphics[scale=0.7]{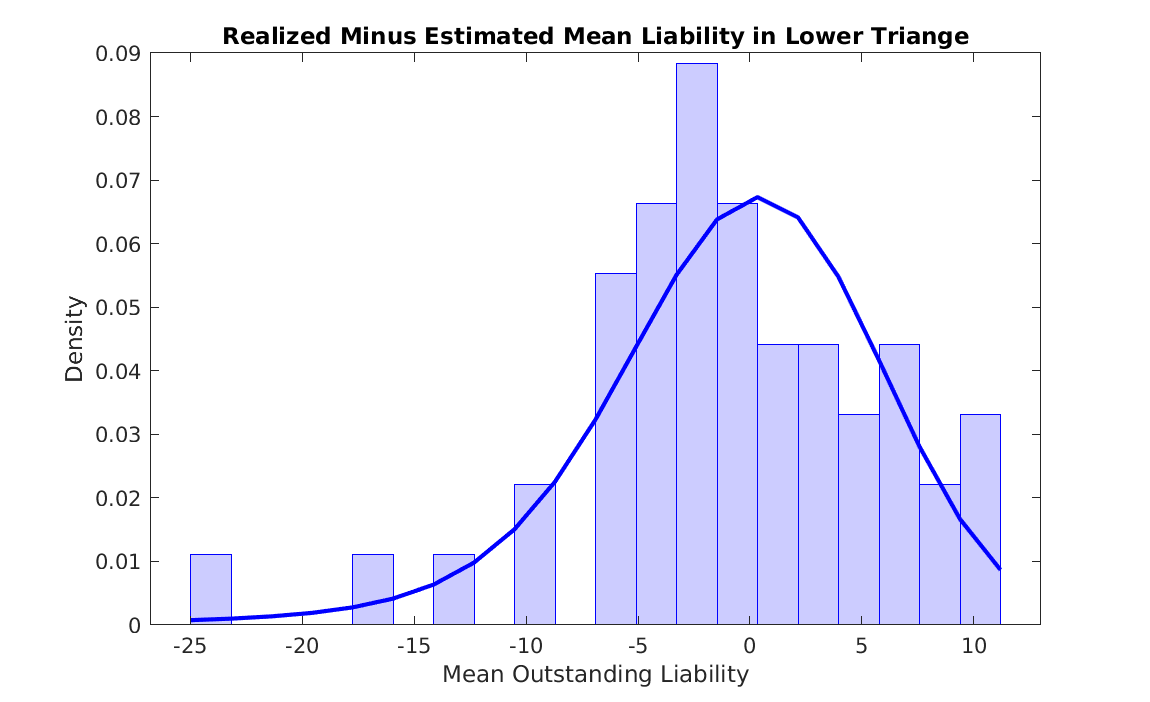}
  \caption{Mean outstanding losses calculated from Table \ref{tab:Stable 1 LoB table} minus the realized losses in the 50 simulated triangles, fitted to a normal distribution (blue line) }
 \label{fig:realized losses}
\end{figure}

In Table \ref{tab:Stable 1 LoB table}, the estimates for $\eta$ are slightly biased. This is in turn reflected in a small underestimation of the outstanding losses in Figure \ref{fig:realized losses}. We suspect this is due to the fact that compared to the MLEs, there is an extra parameter to optimize: the previously discussed regularization parameter $\lambda$. We conduct a crude search by checking which values of $\lambda$ give the best performance in a small series of simulations. A more systematic treatment could potentially offer better results and eliminate any bias. As previously mentioned, see discussion in \cite{kotchoni2012applications} for more on treatment of $\lambda$.

\begin{table}[h!]
\centering
\begin{tabular}{cccccccccccccc}
\cline{3-6} \cline{10-13}
                                  & \multicolumn{1}{c|}{}     & \multicolumn{2}{c|}{CGMM}                               & \multicolumn{2}{c|}{Chain Ladder}                       &                       &                                  & \multicolumn{1}{c|}{}     & \multicolumn{2}{c|}{CGMM}                               & \multicolumn{2}{c|}{MLE}                                &  \\ \cline{2-6} \cline{9-13}
\multicolumn{1}{c|}{}             & \multicolumn{1}{c|}{True} & \multicolumn{1}{c|}{Median} & \multicolumn{1}{c|}{SD}   & \multicolumn{1}{c|}{Median} & \multicolumn{1}{c|}{SD}   &                       & \multicolumn{1}{c|}{}            & \multicolumn{1}{c|}{True} & \multicolumn{1}{c|}{Median} & \multicolumn{1}{c|}{SD}   & \multicolumn{1}{c|}{Median} & \multicolumn{1}{c|}{SD}   &  \\ \cline{1-6} \cline{8-13}
\multicolumn{1}{|c|}{$\nu_1$}     & \multicolumn{1}{c|}{5.00} & \multicolumn{1}{c|}{4.71}   & \multicolumn{1}{c|}{0.77} & \multicolumn{1}{c|}{5.12}   & \multicolumn{1}{c|}{1.14} & \multicolumn{1}{c|}{} & \multicolumn{1}{c|}{$\nu_1$}     & \multicolumn{1}{c|}{5.00} & \multicolumn{1}{c|}{4.97}   & \multicolumn{1}{c|}{2.40} & \multicolumn{1}{c|}{4.75}   & \multicolumn{1}{c|}{1.08} &  \\ \cline{1-6} \cline{8-13}
\multicolumn{1}{|c|}{$\nu_2$}     & \multicolumn{1}{c|}{5.00} & \multicolumn{1}{c|}{4.73}   & \multicolumn{1}{c|}{0.59} & \multicolumn{1}{c|}{4.99}   & \multicolumn{1}{c|}{1.24} & \multicolumn{1}{c|}{} & \multicolumn{1}{c|}{$\nu_2$}     & \multicolumn{1}{c|}{5.00} & \multicolumn{1}{c|}{4.59}   & \multicolumn{1}{c|}{2.00} & \multicolumn{1}{c|}{5.24}   & \multicolumn{1}{c|}{1.04} &  \\ \cline{1-6} \cline{8-13}
\multicolumn{1}{|c|}{$\nu_3$}     & \multicolumn{1}{c|}{5.00} & \multicolumn{1}{c|}{4.84}   & \multicolumn{1}{c|}{0.76} & \multicolumn{1}{c|}{5.05}   & \multicolumn{1}{c|}{1.17} & \multicolumn{1}{c|}{} & \multicolumn{1}{c|}{$\nu_3$}     & \multicolumn{1}{c|}{5.00} & \multicolumn{1}{c|}{4.35}   & \multicolumn{1}{c|}{1.67} & \multicolumn{1}{c|}{5.12}   & \multicolumn{1}{c|}{1.10} &  \\ \cline{1-6} \cline{8-13}
\multicolumn{1}{|c|}{$\nu_4$}     & \multicolumn{1}{c|}{5.00} & \multicolumn{1}{c|}{4.64}   & \multicolumn{1}{c|}{0.80} & \multicolumn{1}{c|}{4.95}   & \multicolumn{1}{c|}{1.29} & \multicolumn{1}{c|}{} & \multicolumn{1}{c|}{$\nu_4$}     & \multicolumn{1}{c|}{5.00} & \multicolumn{1}{c|}{4.41}   & \multicolumn{1}{c|}{2.25} & \multicolumn{1}{c|}{5.16}   & \multicolumn{1}{c|}{1.05} &  \\ \cline{1-6} \cline{8-13}
\multicolumn{1}{|c|}{$\nu_5$}     & \multicolumn{1}{c|}{5.00} & \multicolumn{1}{c|}{4.60}   & \multicolumn{1}{c|}{0.66} & \multicolumn{1}{c|}{4.77}   & \multicolumn{1}{c|}{1.39} & \multicolumn{1}{c|}{} & \multicolumn{1}{c|}{$\nu_5$}     & \multicolumn{1}{c|}{5.00} & \multicolumn{1}{c|}{4.39}   & \multicolumn{1}{c|}{2.38} & \multicolumn{1}{c|}{4.90}   & \multicolumn{1}{c|}{1.23} &  \\ \cline{1-6} \cline{8-13}
\multicolumn{1}{|c|}{$\nu_6$}     & \multicolumn{1}{c|}{5.00} & \multicolumn{1}{c|}{4.79}   & \multicolumn{1}{c|}{0.92} & \multicolumn{1}{c|}{5.23}   & \multicolumn{1}{c|}{1.44} & \multicolumn{1}{c|}{} & \multicolumn{1}{c|}{$\nu_6$}     & \multicolumn{1}{c|}{5.00} & \multicolumn{1}{c|}{4.82}   & \multicolumn{1}{c|}{2.42} & \multicolumn{1}{c|}{4.72}   & \multicolumn{1}{c|}{1.20} &  \\ \cline{1-6} \cline{8-13}
\multicolumn{1}{|c|}{$\nu_7$}     & \multicolumn{1}{c|}{5.00} & \multicolumn{1}{c|}{4.49}   & \multicolumn{1}{c|}{0.80} & \multicolumn{1}{c|}{4.77}   & \multicolumn{1}{c|}{1.44} & \multicolumn{1}{c|}{} & \multicolumn{1}{c|}{$\nu_7$}     & \multicolumn{1}{c|}{5.00} & \multicolumn{1}{c|}{4.88}   & \multicolumn{1}{c|}{2.27} & \multicolumn{1}{c|}{5.18}   & \multicolumn{1}{c|}{1.23} &  \\ \cline{1-6} \cline{8-13}
\multicolumn{1}{|c|}{$\nu_8$}     & \multicolumn{1}{c|}{5.00} & \multicolumn{1}{c|}{4.28}   & \multicolumn{1}{c|}{0.75} & \multicolumn{1}{c|}{4.67}   & \multicolumn{1}{c|}{1.14} & \multicolumn{1}{c|}{} & \multicolumn{1}{c|}{$\nu_8$}     & \multicolumn{1}{c|}{5.00} & \multicolumn{1}{c|}{4.29}   & \multicolumn{1}{c|}{2.05} & \multicolumn{1}{c|}{4.57}   & \multicolumn{1}{c|}{1.40} &  \\ \cline{1-6} \cline{8-13}
\multicolumn{1}{|c|}{$\nu_9$}     & \multicolumn{1}{c|}{5.00} & \multicolumn{1}{c|}{4.29}   & \multicolumn{1}{c|}{0.91} & \multicolumn{1}{c|}{4.69}   & \multicolumn{1}{c|}{1.20} & \multicolumn{1}{c|}{} & \multicolumn{1}{c|}{$\nu_9$}     & \multicolumn{1}{c|}{5.00} & \multicolumn{1}{c|}{4.52}   & \multicolumn{1}{c|}{1.75} & \multicolumn{1}{c|}{4.67}   & \multicolumn{1}{c|}{1.42} &  \\ \cline{1-6} \cline{8-13}
\multicolumn{1}{|c|}{$\nu_{10}$}  & \multicolumn{1}{c|}{5.00} & \multicolumn{1}{c|}{4.84}   & \multicolumn{1}{c|}{0.90} & \multicolumn{1}{c|}{5.06}   & \multicolumn{1}{c|}{1.10} & \multicolumn{1}{c|}{} & \multicolumn{1}{c|}{$\nu_{10}$}  & \multicolumn{1}{c|}{5.00} & \multicolumn{1}{c|}{4.64}   & \multicolumn{1}{c|}{1.70} & \multicolumn{1}{c|}{4.78}   & \multicolumn{1}{c|}{1.86} &  \\ \cline{1-6} \cline{8-13}
                                  &                           &                             &                           &                             &                           &                       &                                  &                           &                             &                           &                             &                           &  \\ \cline{1-6} \cline{8-13}
\multicolumn{1}{|c|}{$\eta_1$}    & \multicolumn{1}{c|}{1.00} & \multicolumn{1}{c|}{1.00}   & \multicolumn{1}{c|}{0.00} & \multicolumn{1}{c|}{1.00}   & \multicolumn{1}{c|}{0.00} & \multicolumn{1}{c|}{} & \multicolumn{1}{c|}{$\eta_1$}    & \multicolumn{1}{c|}{1.00} & \multicolumn{1}{c|}{1.00}   & \multicolumn{1}{c|}{0.00} & \multicolumn{1}{c|}{1.00}   & \multicolumn{1}{c|}{0.00} &  \\ \cline{1-6} \cline{8-13}
\multicolumn{1}{|c|}{$\eta_2$}    & \multicolumn{1}{c|}{0.95} & \multicolumn{1}{c|}{0.97}   & \multicolumn{1}{c|}{0.08} & \multicolumn{1}{c|}{0.92}   & \multicolumn{1}{c|}{0.12} & \multicolumn{1}{c|}{} & \multicolumn{1}{c|}{$\eta_2$}    & \multicolumn{1}{c|}{0.95} & \multicolumn{1}{c|}{0.90}   & \multicolumn{1}{c|}{0.12} & \multicolumn{1}{c|}{0.94}   & \multicolumn{1}{c|}{0.15} &  \\ \cline{1-6} \cline{8-13}
\multicolumn{1}{|c|}{$\eta_3$}    & \multicolumn{1}{c|}{0.90} & \multicolumn{1}{c|}{0.88}   & \multicolumn{1}{c|}{0.07} & \multicolumn{1}{c|}{0.87}   & \multicolumn{1}{c|}{0.09} & \multicolumn{1}{c|}{} & \multicolumn{1}{c|}{$\eta_3$}    & \multicolumn{1}{c|}{0.90} & \multicolumn{1}{c|}{0.82}   & \multicolumn{1}{c|}{0.14} & \multicolumn{1}{c|}{0.87}   & \multicolumn{1}{c|}{0.18} &  \\ \cline{1-6} \cline{8-13}
\multicolumn{1}{|c|}{$\eta_4$}    & \multicolumn{1}{c|}{0.85} & \multicolumn{1}{c|}{0.84}   & \multicolumn{1}{c|}{0.09} & \multicolumn{1}{c|}{0.79}   & \multicolumn{1}{c|}{0.11} & \multicolumn{1}{c|}{} & \multicolumn{1}{c|}{$\eta_4$}    & \multicolumn{1}{c|}{0.85} & \multicolumn{1}{c|}{0.76}   & \multicolumn{1}{c|}{0.14} & \multicolumn{1}{c|}{0.79}   & \multicolumn{1}{c|}{0.18} &  \\ \cline{1-6} \cline{8-13}
\multicolumn{1}{|c|}{$\eta_5$}    & \multicolumn{1}{c|}{0.80} & \multicolumn{1}{c|}{0.82}   & \multicolumn{1}{c|}{0.09} & \multicolumn{1}{c|}{0.77}   & \multicolumn{1}{c|}{0.11} & \multicolumn{1}{c|}{} & \multicolumn{1}{c|}{$\eta_5$}    & \multicolumn{1}{c|}{0.80} & \multicolumn{1}{c|}{0.74}   & \multicolumn{1}{c|}{0.14} & \multicolumn{1}{c|}{0.77}   & \multicolumn{1}{c|}{0.18} &  \\ \cline{1-6} \cline{8-13}
\multicolumn{1}{|c|}{$\eta_6$}    & \multicolumn{1}{c|}{0.75} & \multicolumn{1}{c|}{0.76}   & \multicolumn{1}{c|}{0.09} & \multicolumn{1}{c|}{0.74}   & \multicolumn{1}{c|}{0.10} & \multicolumn{1}{c|}{} & \multicolumn{1}{c|}{$\eta_6$}    & \multicolumn{1}{c|}{0.75} & \multicolumn{1}{c|}{0.72}   & \multicolumn{1}{c|}{0.12} & \multicolumn{1}{c|}{0.71}   & \multicolumn{1}{c|}{0.15} &  \\ \cline{1-6} \cline{8-13}
\multicolumn{1}{|c|}{$\eta_7$}    & \multicolumn{1}{c|}{0.70} & \multicolumn{1}{c|}{0.72}   & \multicolumn{1}{c|}{0.10} & \multicolumn{1}{c|}{0.70}   & \multicolumn{1}{c|}{0.12} & \multicolumn{1}{c|}{} & \multicolumn{1}{c|}{$\eta_7$}    & \multicolumn{1}{c|}{0.70} & \multicolumn{1}{c|}{0.67}   & \multicolumn{1}{c|}{0.14} & \multicolumn{1}{c|}{0.66}   & \multicolumn{1}{c|}{0.16} &  \\ \cline{1-6} \cline{8-13}
\multicolumn{1}{|c|}{$\eta_8$}    & \multicolumn{1}{c|}{0.65} & \multicolumn{1}{c|}{0.68}   & \multicolumn{1}{c|}{0.11} & \multicolumn{1}{c|}{0.66}   & \multicolumn{1}{c|}{0.11} & \multicolumn{1}{c|}{} & \multicolumn{1}{c|}{$\eta_8$}    & \multicolumn{1}{c|}{0.65} & \multicolumn{1}{c|}{0.68}   & \multicolumn{1}{c|}{0.17} & \multicolumn{1}{c|}{0.63}   & \multicolumn{1}{c|}{0.22} &  \\ \cline{1-6} \cline{8-13}
\multicolumn{1}{|c|}{$\eta_9$}    & \multicolumn{1}{c|}{0.60} & \multicolumn{1}{c|}{0.66}   & \multicolumn{1}{c|}{0.12} & \multicolumn{1}{c|}{0.58}   & \multicolumn{1}{c|}{0.13} & \multicolumn{1}{c|}{} & \multicolumn{1}{c|}{$\eta_9$}    & \multicolumn{1}{c|}{0.60} & \multicolumn{1}{c|}{0.63}   & \multicolumn{1}{c|}{0.18} & \multicolumn{1}{c|}{0.59}   & \multicolumn{1}{c|}{0.23} &  \\ \cline{1-6} \cline{8-13}
\multicolumn{1}{|c|}{$\eta_{10}$} & \multicolumn{1}{c|}{0.55} & \multicolumn{1}{c|}{0.58}   & \multicolumn{1}{c|}{0.13} & \multicolumn{1}{c|}{0.53}   & \multicolumn{1}{c|}{0.17} & \multicolumn{1}{c|}{} & \multicolumn{1}{c|}{$\eta_{10}$} & \multicolumn{1}{c|}{0.55} & \multicolumn{1}{c|}{0.61}   & \multicolumn{1}{c|}{0.21} & \multicolumn{1}{c|}{0.54}   & \multicolumn{1}{c|}{0.26} &  \\ \cline{1-6} \cline{8-13}
                                  &                           &                             &                           &                             &                           &                       &                                  &                           &                             &                           &                             &                           &  \\ \cline{1-6} \cline{8-13}
\multicolumn{1}{|c|}{$\gamma$}    & \multicolumn{1}{c|}{0.20} & \multicolumn{1}{c|}{0.22}   & \multicolumn{1}{c|}{0.10} & \multicolumn{1}{c|}{0.08}   & \multicolumn{1}{c|}{0.07} & \multicolumn{1}{c|}{} & \multicolumn{1}{c|}{$\gamma$}    & \multicolumn{1}{c|}{0.20} & \multicolumn{1}{c|}{0.19}   & \multicolumn{1}{c|}{0.11} & \multicolumn{1}{c|}{0.13}   & \multicolumn{1}{c|}{0.03} &  \\ \cline{1-6} \cline{8-13}
\end{tabular}
\caption{Statistics of the CGMM estimators in the Tweedie case with $p=1.2$ (overdispersed Poisson, left) and $p=2$ (Gamma, right) after simulating 50 triangles. Included are the chain ladder and MLE estimators for comparison.}
\label{tab:Tweedie 1 LoB table}
\end{table}

\begin{table}[h!]
\centering
\begin{tabular}{cccccccccc}
                                & True                     & Mean                      & SD                        &                       &                                 & True                      & Mean                      & SD                        &  \\ \cline{1-4} \cline{6-9}
\multicolumn{1}{|c|}{$\eta_1$}  & \multicolumn{1}{c|}{5}   & \multicolumn{1}{c|}{4.82} & \multicolumn{1}{c|}{0.30} & \multicolumn{1}{c|}{} & \multicolumn{1}{c|}{$\nu_1$}    & \multicolumn{1}{c|}{1}    & \multicolumn{1}{c|}{1.00} & \multicolumn{1}{c|}{0.00} &  \\ \cline{1-4} \cline{6-9}
\multicolumn{1}{|c|}{$\eta_2$}  & \multicolumn{1}{c|}{5}   & \multicolumn{1}{c|}{4.78} & \multicolumn{1}{c|}{0.28} & \multicolumn{1}{c|}{} & \multicolumn{1}{c|}{$\nu_2$}    & \multicolumn{1}{c|}{0.95} & \multicolumn{1}{c|}{0.96} & \multicolumn{1}{c|}{0.04} &  \\ \cline{1-4} \cline{6-9}
\multicolumn{1}{|c|}{$\eta_3$}  & \multicolumn{1}{c|}{5}   & \multicolumn{1}{c|}{4.80} & \multicolumn{1}{c|}{0.26} & \multicolumn{1}{c|}{} & \multicolumn{1}{c|}{$\nu_3$}    & \multicolumn{1}{c|}{0.9}  & \multicolumn{1}{c|}{0.92} & \multicolumn{1}{c|}{0.04} &  \\ \cline{1-4} \cline{6-9}
\multicolumn{1}{|c|}{$\eta_4$}  & \multicolumn{1}{c|}{5}   & \multicolumn{1}{c|}{4.84} & \multicolumn{1}{c|}{0.25} & \multicolumn{1}{c|}{} & \multicolumn{1}{c|}{$\nu_4$}    & \multicolumn{1}{c|}{0.85} & \multicolumn{1}{c|}{0.86} & \multicolumn{1}{c|}{0.05} &  \\ \cline{1-4} \cline{6-9}
\multicolumn{1}{|c|}{$\eta_5$}  & \multicolumn{1}{c|}{5}   & \multicolumn{1}{c|}{4.83} & \multicolumn{1}{c|}{0.30} & \multicolumn{1}{c|}{} & \multicolumn{1}{c|}{$\nu_5$}    & \multicolumn{1}{c|}{0.8}  & \multicolumn{1}{c|}{0.81} & \multicolumn{1}{c|}{0.05} &  \\ \cline{1-4} \cline{6-9}
\multicolumn{1}{|c|}{$\eta_6$}  & \multicolumn{1}{c|}{5}   & \multicolumn{1}{c|}{4.79} & \multicolumn{1}{c|}{0.23} & \multicolumn{1}{c|}{} & \multicolumn{1}{c|}{$\nu_6$}    & \multicolumn{1}{c|}{0.75} & \multicolumn{1}{c|}{0.75} & \multicolumn{1}{c|}{0.04} &  \\ \cline{1-4} \cline{6-9}
\multicolumn{1}{|c|}{$\eta_7$}  & \multicolumn{1}{c|}{5}   & \multicolumn{1}{c|}{4.80} & \multicolumn{1}{c|}{0.28} & \multicolumn{1}{c|}{} & \multicolumn{1}{c|}{$\nu_7$}    & \multicolumn{1}{c|}{0.7}  & \multicolumn{1}{c|}{0.70} & \multicolumn{1}{c|}{0.05} &  \\ \cline{1-4} \cline{6-9}
\multicolumn{1}{|c|}{$\eta_8$}  & \multicolumn{1}{c|}{5}   & \multicolumn{1}{c|}{4.80} & \multicolumn{1}{c|}{0.27} & \multicolumn{1}{c|}{} & \multicolumn{1}{c|}{$\nu_8$}    & \multicolumn{1}{c|}{0.65} & \multicolumn{1}{c|}{0.66} & \multicolumn{1}{c|}{0.05} &  \\ \cline{1-4} \cline{6-9}
\multicolumn{1}{|c|}{$\eta_9$}  & \multicolumn{1}{c|}{5}   & \multicolumn{1}{c|}{4.84} & \multicolumn{1}{c|}{0.25} & \multicolumn{1}{c|}{} & \multicolumn{1}{c|}{$\nu_9$}    & \multicolumn{1}{c|}{0.6}  & \multicolumn{1}{c|}{0.59} & \multicolumn{1}{c|}{0.06} &  \\ \cline{1-4} \cline{6-9}
\multicolumn{1}{|c|}{$\eta_10$} & \multicolumn{1}{c|}{5}   & \multicolumn{1}{c|}{4.82} & \multicolumn{1}{c|}{0.27} & \multicolumn{1}{c|}{} & \multicolumn{1}{c|}{$\nu_{10}$} & \multicolumn{1}{c|}{0.55} & \multicolumn{1}{c|}{0.54} & \multicolumn{1}{c|}{0.07} &  \\ \cline{1-4} \cline{6-9}
                                &                          &                           &                           &                       &                                 &                           &                           &                           &  \\ \cline{1-4}
\multicolumn{1}{|c|}{$\gamma$}  & \multicolumn{1}{c|}{0.2} & \multicolumn{1}{c|}{0.18} & \multicolumn{1}{c|}{0.02} &                       &                                 &                           &                           &                           &  \\ \cline{1-4}
\end{tabular}
\caption{Statistics of the CGMM estimators in the stable case after simulating 50 triangles}
\label{tab:Stable 1 LoB table}
\end{table}

\subsection{Multiple LoB Illustration}\label{sec: muti lob}

In this section, we extend all the same methods from the single LoB to two LoBs. We simulate 50 joint triangles from a two-dimensional model of the kind specified by (\ref{eq: ABRM vector}). To create our objective we use a moment condition of the form 
\begin{align*}\label{eq: Multi MGF moment fun}
h( (\tau_1,\tau_2), (\theta_1, \theta_2, \theta_Z) : \mathbf{X} ) 
&= e^{ \tau_1 a^{(1)}Y^{(1)} + \tau_2 a^{(2)}Y^{(2)} + (\tau_1+\tau_2) b Z } \\ 
&\quad - M_{Y^{(1)}|\theta_1}(\tau_1 a^{(1)})M_{Y^{(2)}|\theta_2}(\tau_2 a^{(2)})M_{Z|\theta_Z}( (\tau_1+\tau_2)b ) .
\end{align*}

For the Stable case, we once again consider evaluating using $-X$ and the half-plane containing $\tau_i>0$, but a practical problem immediately presents itself. As alluded to in Section \ref{sec: numerical CGMM}, $\mathcal{K}^{-1/2}h$ produces an objective function with little curvature around the optimum, being basically flat for much of its support. Our current workaround to produce the results in Table \ref{tab:Multi Stab Results} is as follows. First, we make use of a change of variable to concentrate as many of the quadrature points as possible where the function has the most substantial value. Second, we multiply the corresponding objective by a large number. This guarantees that any optimizing algorithm unable to distinguish between a flat region and a very lightly curved one would not terminate prematurely. That being said, formally understanding the relationship between moment conditions and the resulting objective, along with a more systematic approach to numerical optimization, remains an avenue for future work.

The results here are not as precise as in the previous section. However, virtually all parameters are within a standard deviation. Importantly, the CGMM is able to capture the systematic parameters motivating the exercise in the first place. 

It goes without saying that the results for the single LoB cases seem more accurate. One aspect of the CGMM we lament is the degree of subjectivity of choices in the methodology. For example, to guarantee that the regularization term and MSE in Eq.\ (\ref{eq:regularization}) are balanced (guaranteeing that we converge to the ``correct'' solution as $\lambda \rightarrow 0$), we need to standardize the columns of $\mathcal{K}$. In the multiple LoB case, $\mathcal{K}$ is rather sparse and hence very sensitive to the choice of standardization, dramatically affecting the results. This is an example of the kind of problem we encounter in the multiple LoB case that is either not present or at least not as important as in the single LoB case. Furthermore, practical results (especially vs the MLE) would still take on the order of hours or days to produce, making thorough experimentation slow. That said, we are confident that finding the proper regularization is simply a matter of trial and error. 

In the next section, we apply the same method to a well-studied data set, and succeed in producing seemingly favourable results. This suggests even as a preliminary conclusion that the CGMM can compete with existing methods. 

\begin{table}[h!]
\centering
\begin{tabular}{ccccccccc}
\cline{1-4} \cline{6-9}
\multicolumn{4}{|c|}{Line 1}                                                                                                & \multicolumn{1}{c|}{} & \multicolumn{4}{c|}{Line 2}                                                                                                \\ \cline{1-4} \cline{6-9} 
\multicolumn{1}{|c|}{Parameter}       & \multicolumn{1}{c|}{True} & \multicolumn{1}{c|}{Median} & \multicolumn{1}{c|}{SD}   & \multicolumn{1}{c|}{} & \multicolumn{1}{c|}{Parameter}       & \multicolumn{1}{c|}{True} & \multicolumn{1}{c|}{Median} & \multicolumn{1}{c|}{SD}   \\ \cline{1-4} \cline{6-9} 
\multicolumn{1}{|c|}{$\eta_{1}^{1}$}  & \multicolumn{1}{c|}{5.00} & \multicolumn{1}{c|}{4.42}   & \multicolumn{1}{c|}{0.35} & \multicolumn{1}{c|}{} & \multicolumn{1}{c|}{$\eta_{1}^{2}$}  & \multicolumn{1}{c|}{4.00} & \multicolumn{1}{c|}{3.78}   & \multicolumn{1}{c|}{0.52} \\ \cline{1-4} \cline{6-9} 
\multicolumn{1}{|c|}{$\eta_{2}^{1}$}  & \multicolumn{1}{c|}{5.00} & \multicolumn{1}{c|}{4.48}   & \multicolumn{1}{c|}{0.46} & \multicolumn{1}{c|}{} & \multicolumn{1}{c|}{$\eta_{2}^{2}$}  & \multicolumn{1}{c|}{4.00} & \multicolumn{1}{c|}{3.68}   & \multicolumn{1}{c|}{0.44} \\ \cline{1-4} \cline{6-9} 
\multicolumn{1}{|c|}{$\eta_{3}^{1}$}  & \multicolumn{1}{c|}{5.00} & \multicolumn{1}{c|}{4.44}   & \multicolumn{1}{c|}{0.35} & \multicolumn{1}{c|}{} & \multicolumn{1}{c|}{$\eta_{3}^{2}$}  & \multicolumn{1}{c|}{4.00} & \multicolumn{1}{c|}{3.80}   & \multicolumn{1}{c|}{0.52} \\ \cline{1-4} \cline{6-9} 
\multicolumn{1}{|c|}{$\eta_{4}^{1}$}  & \multicolumn{1}{c|}{5.00} & \multicolumn{1}{c|}{4.40}   & \multicolumn{1}{c|}{0.39} & \multicolumn{1}{c|}{} & \multicolumn{1}{c|}{$\eta_{4}^{2}$}  & \multicolumn{1}{c|}{4.00} & \multicolumn{1}{c|}{3.94}   & \multicolumn{1}{c|}{0.47} \\ \cline{1-4} \cline{6-9} 
\multicolumn{1}{|c|}{$\eta_{5}^{1}$}  & \multicolumn{1}{c|}{5.00} & \multicolumn{1}{c|}{4.48}   & \multicolumn{1}{c|}{0.60} & \multicolumn{1}{c|}{} & \multicolumn{1}{c|}{$\eta_{5}^{2}$}  & \multicolumn{1}{c|}{4.00} & \multicolumn{1}{c|}{3.89}   & \multicolumn{1}{c|}{0.52} \\ \cline{1-4} \cline{6-9} 
\multicolumn{1}{|c|}{$\eta_{6}^{1}$}  & \multicolumn{1}{c|}{5.00} & \multicolumn{1}{c|}{4.41}   & \multicolumn{1}{c|}{0.47} & \multicolumn{1}{c|}{} & \multicolumn{1}{c|}{$\eta_{6}^{2}$}  & \multicolumn{1}{c|}{4.00} & \multicolumn{1}{c|}{3.89}   & \multicolumn{1}{c|}{0.40} \\ \cline{1-4} \cline{6-9} 
\multicolumn{1}{|c|}{$\eta_{7}^{1}$}  & \multicolumn{1}{c|}{5.00} & \multicolumn{1}{c|}{4.43}   & \multicolumn{1}{c|}{0.40} & \multicolumn{1}{c|}{} & \multicolumn{1}{c|}{$\eta_{7}^{2}$}  & \multicolumn{1}{c|}{4.00} & \multicolumn{1}{c|}{3.85}   & \multicolumn{1}{c|}{0.39} \\ \cline{1-4} \cline{6-9} 
\multicolumn{1}{|c|}{$\eta_{8}^{1}$}  & \multicolumn{1}{c|}{5.00} & \multicolumn{1}{c|}{4.44}   & \multicolumn{1}{c|}{0.39} & \multicolumn{1}{c|}{} & \multicolumn{1}{c|}{$\eta_{8}^{2}$}  & \multicolumn{1}{c|}{4.00} & \multicolumn{1}{c|}{3.98}   & \multicolumn{1}{c|}{0.40} \\ \cline{1-4} \cline{6-9} 
\multicolumn{1}{|c|}{$\eta_{9}^{1}$}  & \multicolumn{1}{c|}{5.00} & \multicolumn{1}{c|}{4.55}   & \multicolumn{1}{c|}{0.63} & \multicolumn{1}{c|}{} & \multicolumn{1}{c|}{$\eta_{9}^{2}$}  & \multicolumn{1}{c|}{4.00} & \multicolumn{1}{c|}{3.92}   & \multicolumn{1}{c|}{0.85} \\ \cline{1-4} \cline{6-9} 
\multicolumn{1}{|c|}{$\eta_{10}^{1}$} & \multicolumn{1}{c|}{5.00} & \multicolumn{1}{c|}{4.66}   & \multicolumn{1}{c|}{0.31} & \multicolumn{1}{c|}{} & \multicolumn{1}{c|}{$\eta_{10}^{2}$} & \multicolumn{1}{c|}{4.00} & \multicolumn{1}{c|}{4.02}   & \multicolumn{1}{c|}{0.43} \\ \cline{1-4} \cline{6-9} 
                                      &                           &                             &                           &                       &                                      &                           &                             &                           \\ \cline{1-4} \cline{6-9} 
\multicolumn{1}{|c|}{$\nu_{1}^{1}$}   & \multicolumn{1}{c|}{1.00} & \multicolumn{1}{c|}{1.00}   & \multicolumn{1}{c|}{0.00} & \multicolumn{1}{c|}{} & \multicolumn{1}{c|}{$\nu_{1}^{2}$}   & \multicolumn{1}{c|}{1.00} & \multicolumn{1}{c|}{1.00}   & \multicolumn{1}{c|}{0.00} \\ \cline{1-4} \cline{6-9} 
\multicolumn{1}{|c|}{$\nu_{2}^{1}$}   & \multicolumn{1}{c|}{0.93} & \multicolumn{1}{c|}{1.01}   & \multicolumn{1}{c|}{0.08} & \multicolumn{1}{c|}{} & \multicolumn{1}{c|}{$\nu_{2}^{2}$}   & \multicolumn{1}{c|}{0.95} & \multicolumn{1}{c|}{0.96}   & \multicolumn{1}{c|}{0.09} \\ \cline{1-4} \cline{6-9} 
\multicolumn{1}{|c|}{$\nu_{3}^{1}$}   & \multicolumn{1}{c|}{0.87} & \multicolumn{1}{c|}{0.95}   & \multicolumn{1}{c|}{0.07} & \multicolumn{1}{c|}{} & \multicolumn{1}{c|}{$\nu_{3}^{2}$}   & \multicolumn{1}{c|}{0.90} & \multicolumn{1}{c|}{0.92}   & \multicolumn{1}{c|}{0.10} \\ \cline{1-4} \cline{6-9} 
\multicolumn{1}{|c|}{$\nu_{4}^{1}$}   & \multicolumn{1}{c|}{0.80} & \multicolumn{1}{c|}{0.88}   & \multicolumn{1}{c|}{0.07} & \multicolumn{1}{c|}{} & \multicolumn{1}{c|}{$\nu_{4}^{2}$}   & \multicolumn{1}{c|}{0.85} & \multicolumn{1}{c|}{0.85}   & \multicolumn{1}{c|}{0.11} \\ \cline{1-4} \cline{6-9} 
\multicolumn{1}{|c|}{$\nu_{5}^{1}$}   & \multicolumn{1}{c|}{0.73} & \multicolumn{1}{c|}{0.81}   & \multicolumn{1}{c|}{0.08} & \multicolumn{1}{c|}{} & \multicolumn{1}{c|}{$\nu_{5}^{2}$}   & \multicolumn{1}{c|}{0.80} & \multicolumn{1}{c|}{0.81}   & \multicolumn{1}{c|}{0.08} \\ \cline{1-4} \cline{6-9} 
\multicolumn{1}{|c|}{$\nu_{6}^{1}$}   & \multicolumn{1}{c|}{0.67} & \multicolumn{1}{c|}{0.73}   & \multicolumn{1}{c|}{0.06} & \multicolumn{1}{c|}{} & \multicolumn{1}{c|}{$\nu_{6}^{2}$}   & \multicolumn{1}{c|}{0.75} & \multicolumn{1}{c|}{0.76}   & \multicolumn{1}{c|}{0.09} \\ \cline{1-4} \cline{6-9} 
\multicolumn{1}{|c|}{$\nu_{7}^{1}$}   & \multicolumn{1}{c|}{0.60} & \multicolumn{1}{c|}{0.65}   & \multicolumn{1}{c|}{0.05} & \multicolumn{1}{c|}{} & \multicolumn{1}{c|}{$\nu_{7}^{2}$}   & \multicolumn{1}{c|}{0.70} & \multicolumn{1}{c|}{0.69}   & \multicolumn{1}{c|}{0.09} \\ \cline{1-4} \cline{6-9} 
\multicolumn{1}{|c|}{$\nu_{8}^{1}$}   & \multicolumn{1}{c|}{0.53} & \multicolumn{1}{c|}{0.60}   & \multicolumn{1}{c|}{0.08} & \multicolumn{1}{c|}{} & \multicolumn{1}{c|}{$\nu_{8}^{2}$}   & \multicolumn{1}{c|}{0.65} & \multicolumn{1}{c|}{0.65}   & \multicolumn{1}{c|}{0.11} \\ \cline{1-4} \cline{6-9} 
\multicolumn{1}{|c|}{$\nu_{9}^{1}$}   & \multicolumn{1}{c|}{0.47} & \multicolumn{1}{c|}{0.51}   & \multicolumn{1}{c|}{0.06} & \multicolumn{1}{c|}{} & \multicolumn{1}{c|}{$\nu_{9}^{2}$}   & \multicolumn{1}{c|}{0.60} & \multicolumn{1}{c|}{0.56}   & \multicolumn{1}{c|}{0.10} \\ \cline{1-4} \cline{6-9} 
\multicolumn{1}{|c|}{$\nu_{10}^{1}$}  & \multicolumn{1}{c|}{0.40} & \multicolumn{1}{c|}{0.43}   & \multicolumn{1}{c|}{0.17} & \multicolumn{1}{c|}{} & \multicolumn{1}{c|}{$\nu_{10}^{2}$}  & \multicolumn{1}{c|}{0.55} & \multicolumn{1}{c|}{0.54}   & \multicolumn{1}{c|}{0.22} \\ \cline{1-4} \cline{6-9} 
                                      &                           &                             &                           &                       &                                      &                           &                             &                           \\ \cline{1-4} \cline{6-9} 
\multicolumn{1}{|c|}{$\gamma^{1}$}    & \multicolumn{1}{c|}{0.20} & \multicolumn{1}{c|}{0.16}   & \multicolumn{1}{c|}{0.10} & \multicolumn{1}{c|}{} & \multicolumn{1}{c|}{$\gamma^{2}$}    & \multicolumn{1}{c|}{0.30} & \multicolumn{1}{c|}{0.15}   & \multicolumn{1}{c|}{0.09} \\ \cline{1-4} \cline{6-9} 
                                      &                           &                             &                           &                       &                                      &                           &                             &                           \\ \cline{1-4}
\multicolumn{4}{|c|}{Systematic Parameters}                                                                                 &                       &                                      &                           &                             &                           \\ \cline{1-4}
\multicolumn{1}{|c|}{Parameter}       & \multicolumn{1}{c|}{1.00} & \multicolumn{1}{c|}{Median} & \multicolumn{1}{c|}{SD}   &                       &                                      &                           &                             &                           \\ \cline{1-4}
\multicolumn{1}{|c|}{$\mu$}           & \multicolumn{1}{c|}{0.10} & \multicolumn{1}{c|}{0.17}   & \multicolumn{1}{c|}{0.12} &                       &                                      &                           &                             &                           \\ \cline{1-4}
\multicolumn{1}{|c|}{$\sigma$}        & \multicolumn{1}{c|}{0.10} & \multicolumn{1}{c|}{0.12}   & \multicolumn{1}{c|}{0.07} &                       &                                      &                           &                             &                           \\ \cline{1-4}
\end{tabular}
\caption{The results of CGMM estimation from 50 joint pairs of triangles}
\label{tab:Multi Stab Results}
\end{table}

\newpage

\section{Real-world Data Analysis}\label{sec: RealLoss}

In this section, we use the data first used in \cite{zhang2013predicting} and \cite{avanzi2016stochastic} from the Pennsylvania National Insurance Group (Schedule P, see Table \ref{AutoData1} and \ref{AutoData2}) to study multivariate and copula Tweedie-based loss models from a Bayesian perspective. We estimate every model parameter using our multivariate CGMM methodology except for the Tweedie power parameter $p$, which we set \textit{a priori} at $p=1.32$. This is the value found in \cite{avanzi2016stochastic}, derived from an analysis of the likelihood. 

One advantage of the Bayesian methods employed in the aforementioned works is that they come equipped with a built-in uncertainty estimate for a single loss-reserve triangle. Calculating the variance of model parameters and confidence intervals is quite laborious in the CGMM. We instead opt to use the parametric bootstrap described in \cite{wuthrich2008stochastic}. To briefly summarize, we estimate the model parameters from the Schedule P data and generate new loss triangles from the results, re-estimating the parameters in this bootstrapped sample and outstanding claims. In Table \ref{tab:MultiTweed Results}, we can see the summary statistics from the bootstrapped samples, with the resulting outstanding claims reserve statistics in \ref{tab:cumsum outstanding}. We find that the results of Table \ref{tab:MultiTweed Results} are similar to the celebrated chain ladder estimates for $\eta$ and $\nu$. Given that the Tweedie is in some sense a parametric equivalent to these classic estimators, this is both unsurprising and also a validation of the CGMM estimates. 

Table \ref{tab:comparison} summarizes our outstanding estimates alongside previous results using the same data set. There appear to be smaller variances in the CGMM results. A possible reason is the fact that in other methods relatively harsh uniform priors were used in a Bayesian framework; besides, the \cite{zhang2013predicting} results use heavier-tailed log-Normal marginals. In light of the simulation results in the previous section, we must also consider the possibility that the CGMM may be systematically underestimating the variance of outstanding claims.

\begin{figure}[h!]
  \centering
  \includegraphics[scale=0.7]{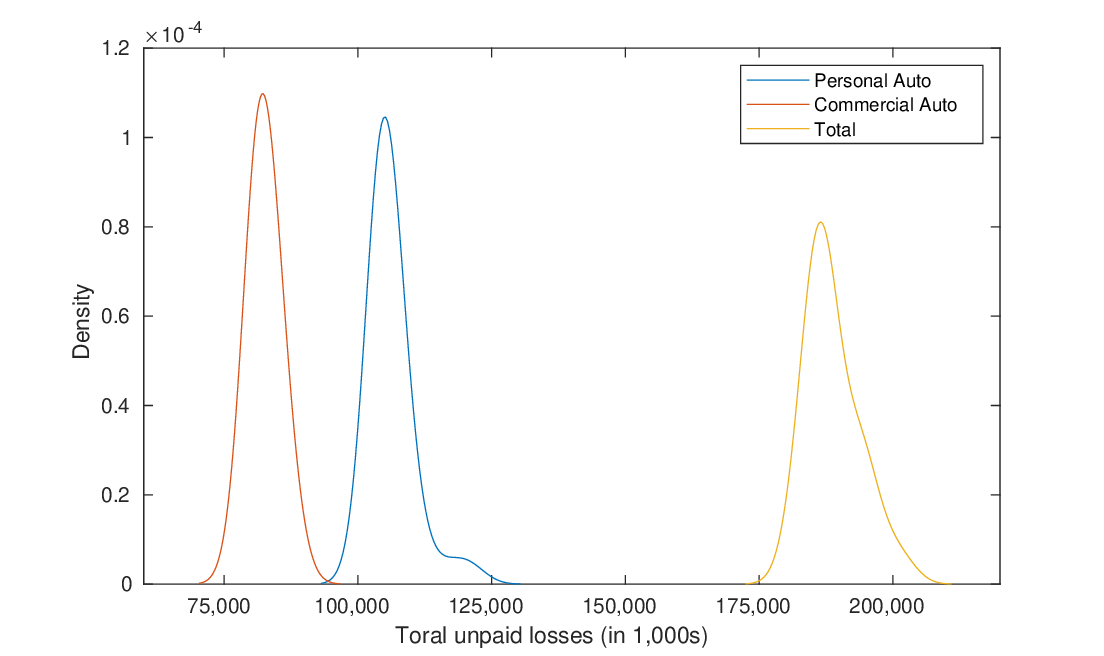}
  \caption{ Kernel density estimates for  outstanding loss reserves, calculated via parametric bootstrap }
 \label{fig:}
\end{figure}

\begin{table}[h!]
\resizebox{\columnwidth}{!}{%
\centering 
\begin{tabular}{l|r|r|r|r|r|r|r|r|r|}
\cline{2-10}
                                         & \multicolumn{3}{c|}{Personal}                                                        & \multicolumn{3}{c|}{Commercial}                                                      & \multicolumn{3}{c|}{Total}                                                           \\ \hline
\multicolumn{1}{|c|}{Model}              & \multicolumn{1}{c|}{Median} & \multicolumn{1}{c|}{SD} & \multicolumn{1}{c|}{Q(0.99)} & \multicolumn{1}{c|}{Median} & \multicolumn{1}{c|}{SD} & \multicolumn{1}{c|}{Q(0.99)} & \multicolumn{1}{c|}{Median} & \multicolumn{1}{c|}{SD} & \multicolumn{1}{c|}{Q(0.99)} \\ \hline
\multicolumn{1}{|l|}{Multivariate Tweedie (CGMM)}  & 104,935                     & 3,750                   & 121,377                      & 82,038                      & 2,052                   & 87,715                       & 187,542                     & 4,786                   & 201,928                      \\ \hline
\multicolumn{1}{|l|}{Multivariate Tweedie (Bayes)} & 103,374                     & 9,373                   & 127,075                      & 88,385                      & 9,029                   & 112,258                      & 192,148                     & 13,780                  & 226,110                      \\ \hline
\multicolumn{1}{|l|}{Clayton Copula Tweedie (Bayes)}            & 103,674                     & 18,742                  & 166,187                      & 91,067                      & 15,820                  & 135,924                      & 194,741                     & 28,376                  & 283,931                      \\ \hline
\multicolumn{1}{|l|}{Gaussian Copula Tweedie (Bayes)}           & 107,930                     & 21,502                  & 172,161                      & 92,773                      & 17,902                  & 147,734                      & 200,703                     & 31,333                  & 295,900                      \\ \hline
\end{tabular}
}
\caption{Comparison of estimated outstanding reserves from our work (CGMM), \cite{avanzi2016stochastic} (Bayesian estimation applied to the multivariate Tweedie), and \cite{zhang2013predicting} (Bayesian estimation of a model with Tweedie marginals and copula dependence) }\label{tab:comparison}
\end{table}

\begin{table}[h!]
\centering 
\begin{tabular}{ccccclccccc}
\multicolumn{5}{c}{Personal Auto}                                                                                                                                   &                       & \multicolumn{5}{c}{Commercial Auto}                                                                                                                                \\ \cline{2-5} \cline{8-11} 
\multicolumn{1}{c|}{}                   & \multicolumn{1}{c|}{Median} & \multicolumn{1}{c|}{SD} & \multicolumn{1}{c|}{Q(0.05)} & \multicolumn{1}{c|}{Q(0.95)} &                       & \multicolumn{1}{c|}{}                  & \multicolumn{1}{c|}{Median} & \multicolumn{1}{c|}{SD} & \multicolumn{1}{c|}{Q(0.05)} & \multicolumn{1}{c|}{Q(0.95)} \\ \cline{1-5} \cline{7-11} 
\multicolumn{1}{|c|}{$\eta_1^{(1)}$}    & \multicolumn{1}{c|}{0.2635} & \multicolumn{1}{c|}{0.0071}   & \multicolumn{1}{c|}{0.2506}  & \multicolumn{1}{c|}{0.2706}  & \multicolumn{1}{l|}{} & \multicolumn{1}{c|}{$\eta_1^{(2)}$}    & \multicolumn{1}{c|}{0.1478} & \multicolumn{1}{c|}{0.0142}   & \multicolumn{1}{c|}{0.1264}  & \multicolumn{1}{c|}{0.1675}  \\ \cline{1-5} \cline{7-11} 
\multicolumn{1}{|c|}{$\eta_2^{(1)}$}    & \multicolumn{1}{c|}{0.2429} & \multicolumn{1}{c|}{0.0067}   & \multicolumn{1}{c|}{0.2328}  & \multicolumn{1}{c|}{0.2512}  & \multicolumn{1}{l|}{} & \multicolumn{1}{c|}{$\eta_2^{(2)}$}    & \multicolumn{1}{c|}{0.1528} & \multicolumn{1}{c|}{0.0107}   & \multicolumn{1}{c|}{0.1335}  & \multicolumn{1}{c|}{0.1653}  \\ \cline{1-5} \cline{7-11} 
\multicolumn{1}{|c|}{$\eta_3^{(1)}$}    & \multicolumn{1}{c|}{0.2295} & \multicolumn{1}{c|}{0.0058}   & \multicolumn{1}{c|}{0.2250}  & \multicolumn{1}{c|}{0.2430}  & \multicolumn{1}{l|}{} & \multicolumn{1}{c|}{$\eta_3^{(2)}$}    & \multicolumn{1}{c|}{0.1517} & \multicolumn{1}{c|}{0.0148}   & \multicolumn{1}{c|}{0.1291}  & \multicolumn{1}{c|}{0.1687}  \\ \cline{1-5} \cline{7-11} 
\multicolumn{1}{|c|}{$\eta_4^{(1)}$}    & \multicolumn{1}{c|}{0.2299} & \multicolumn{1}{c|}{0.0082}   & \multicolumn{1}{c|}{0.2233}  & \multicolumn{1}{c|}{0.2516}  & \multicolumn{1}{l|}{} & \multicolumn{1}{c|}{$\eta_4^{(2)}$}    & \multicolumn{1}{c|}{0.1432} & \multicolumn{1}{c|}{0.0180}   & \multicolumn{1}{c|}{0.1203}  & \multicolumn{1}{c|}{0.1731}  \\ \cline{1-5} \cline{7-11} 
\multicolumn{1}{|c|}{$\eta_5^{(1)}$}    & \multicolumn{1}{c|}{0.2324} & \multicolumn{1}{c|}{0.0062}   & \multicolumn{1}{c|}{0.2271}  & \multicolumn{1}{c|}{0.2457}  & \multicolumn{1}{l|}{} & \multicolumn{1}{c|}{$\eta_5^{(2)}$}    & \multicolumn{1}{c|}{0.1611} & \multicolumn{1}{c|}{0.0200}   & \multicolumn{1}{c|}{0.1286}  & \multicolumn{1}{c|}{0.1830}  \\ \cline{1-5} \cline{7-11} 
\multicolumn{1}{|c|}{$\eta_6^{(1)}$}    & \multicolumn{1}{c|}{0.2268} & \multicolumn{1}{c|}{0.0108}   & \multicolumn{1}{c|}{0.2191}  & \multicolumn{1}{c|}{0.2424}  & \multicolumn{1}{l|}{} & \multicolumn{1}{c|}{$\eta_6^{(2)}$}    & \multicolumn{1}{c|}{0.1457} & \multicolumn{1}{c|}{0.0138}   & \multicolumn{1}{c|}{0.1327}  & \multicolumn{1}{c|}{0.1797}  \\ \cline{1-5} \cline{7-11} 
\multicolumn{1}{|c|}{$\eta_7^{(1)}$}    & \multicolumn{1}{c|}{0.2453} & \multicolumn{1}{c|}{0.0042}   & \multicolumn{1}{c|}{0.2385}  & \multicolumn{1}{c|}{0.2527}  & \multicolumn{1}{l|}{} & \multicolumn{1}{c|}{$\eta_7^{(2)}$}    & \multicolumn{1}{c|}{0.1804} & \multicolumn{1}{c|}{0.0105}   & \multicolumn{1}{c|}{0.1649}  & \multicolumn{1}{c|}{0.1919}  \\ \cline{1-5} \cline{7-11} 
\multicolumn{1}{|c|}{$\eta_8^{(1)}$}    & \multicolumn{1}{c|}{0.2495} & \multicolumn{1}{c|}{0.0042}   & \multicolumn{1}{c|}{0.2400}  & \multicolumn{1}{c|}{0.2526}  & \multicolumn{1}{l|}{} & \multicolumn{1}{c|}{$\eta_8^{(2)}$}    & \multicolumn{1}{c|}{0.1844} & \multicolumn{1}{c|}{0.0084}   & \multicolumn{1}{c|}{0.1655}  & \multicolumn{1}{c|}{0.1930}  \\ \cline{1-5} \cline{7-11} 
\multicolumn{1}{|c|}{$\eta_9^{(1)}$}    & \multicolumn{1}{c|}{0.2946} & \multicolumn{1}{c|}{0.0068}   & \multicolumn{1}{c|}{0.2887}  & \multicolumn{1}{c|}{0.3008}  & \multicolumn{1}{l|}{} & \multicolumn{1}{c|}{$\eta_9^{(2)}$}    & \multicolumn{1}{c|}{0.1982} & \multicolumn{1}{c|}{0.0054}   & \multicolumn{1}{c|}{0.1900}  & \multicolumn{1}{c|}{0.2033}  \\ \cline{1-5} \cline{7-11} 
\multicolumn{1}{|c|}{$\eta_{10}^{(1)}$} & \multicolumn{1}{c|}{0.2792} & \multicolumn{1}{c|}{0.0018}   & \multicolumn{1}{c|}{0.2773}  & \multicolumn{1}{c|}{0.2815}  & \multicolumn{1}{l|}{} & \multicolumn{1}{c|}{$\eta_{10}^{(2)}$} & \multicolumn{1}{c|}{0.2107} & \multicolumn{1}{c|}{0.0025}   & \multicolumn{1}{c|}{0.2061}  & \multicolumn{1}{c|}{0.2140}  \\ \cline{1-5} \cline{7-11} 
\multicolumn{1}{l}{}                    & \multicolumn{1}{l}{}        & \multicolumn{1}{l}{}          & \multicolumn{1}{l}{}         & \multicolumn{1}{l}{}         &                       & \multicolumn{1}{l}{}                   & \multicolumn{1}{l}{}        & \multicolumn{1}{l}{}          & \multicolumn{1}{l}{}         & \multicolumn{1}{l}{}         \\ \cline{1-5} \cline{7-11} 
\multicolumn{1}{|c|}{$\nu_1^{(1)}$}     & \multicolumn{1}{c|}{1.0000} & \multicolumn{1}{c|}{0.0000}   & \multicolumn{1}{c|}{1.0000}  & \multicolumn{1}{c|}{1.0000}  & \multicolumn{1}{l|}{} & \multicolumn{1}{c|}{$\nu_1^{(2)}$}     & \multicolumn{1}{c|}{1.0000} & \multicolumn{1}{c|}{0.0000}   & \multicolumn{1}{c|}{1.0000}  & \multicolumn{1}{c|}{1.0000}  \\ \cline{1-5} \cline{7-11} 
\multicolumn{1}{|c|}{$\nu_2^{(1)}$}     & \multicolumn{1}{c|}{0.9951} & \multicolumn{1}{c|}{0.0044}   & \multicolumn{1}{c|}{0.9896}  & \multicolumn{1}{c|}{0.9999}  & \multicolumn{1}{l|}{} & \multicolumn{1}{c|}{$\nu_2^{(2)}$}     & \multicolumn{1}{c|}{0.9999} & \multicolumn{1}{c|}{0.0023}   & \multicolumn{1}{c|}{0.9944}  & \multicolumn{1}{c|}{1.0000}  \\ \cline{1-5} \cline{7-11} 
\multicolumn{1}{|c|}{$\nu_3^{(1)}$}     & \multicolumn{1}{c|}{0.5828} & \multicolumn{1}{c|}{0.0045}   & \multicolumn{1}{c|}{0.5787}  & \multicolumn{1}{c|}{0.5883}  & \multicolumn{1}{l|}{} & \multicolumn{1}{c|}{$\nu_3^{(2)}$}     & \multicolumn{1}{c|}{0.8753} & \multicolumn{1}{c|}{0.0036}   & \multicolumn{1}{c|}{0.8685}  & \multicolumn{1}{c|}{0.8770}  \\ \cline{1-5} \cline{7-11} 
\multicolumn{1}{|c|}{$\nu_4^{(1)}$}     & \multicolumn{1}{c|}{0.3518} & \multicolumn{1}{c|}{0.0100}   & \multicolumn{1}{c|}{0.3393}  & \multicolumn{1}{c|}{0.3691}  & \multicolumn{1}{l|}{} & \multicolumn{1}{c|}{$\nu_4^{(2)}$}     & \multicolumn{1}{c|}{0.7074} & \multicolumn{1}{c|}{0.0044}   & \multicolumn{1}{c|}{0.7036}  & \multicolumn{1}{c|}{0.7220}  \\ \cline{1-5} \cline{7-11} 
\multicolumn{1}{|c|}{$\nu_5^{(1)}$}     & \multicolumn{1}{c|}{0.1977} & \multicolumn{1}{c|}{0.0139}   & \multicolumn{1}{c|}{0.1802}  & \multicolumn{1}{c|}{0.2194}  & \multicolumn{1}{l|}{} & \multicolumn{1}{c|}{$\nu_5^{(2)}$}     & \multicolumn{1}{c|}{0.4635} & \multicolumn{1}{c|}{0.0075}   & \multicolumn{1}{c|}{0.4578}  & \multicolumn{1}{c|}{0.4744}  \\ \cline{1-5} \cline{7-11} 
\multicolumn{1}{|c|}{$\nu_6^{(1)}$}     & \multicolumn{1}{c|}{0.0932} & \multicolumn{1}{c|}{0.0100}   & \multicolumn{1}{c|}{0.0839}  & \multicolumn{1}{c|}{0.1089}  & \multicolumn{1}{l|}{} & \multicolumn{1}{c|}{$\nu_6^{(2)}$}     & \multicolumn{1}{c|}{0.1915} & \multicolumn{1}{c|}{0.0111}   & \multicolumn{1}{c|}{0.1875}  & \multicolumn{1}{c|}{0.2135}  \\ \cline{1-5} \cline{7-11} 
\multicolumn{1}{|c|}{$\nu_7^{(1)}$}     & \multicolumn{1}{c|}{0.0310} & \multicolumn{1}{c|}{0.0038}   & \multicolumn{1}{c|}{0.0223}  & \multicolumn{1}{c|}{0.0333}  & \multicolumn{1}{l|}{} & \multicolumn{1}{c|}{$\nu_7^{(2)}$}     & \multicolumn{1}{c|}{0.1428} & \multicolumn{1}{c|}{0.0045}   & \multicolumn{1}{c|}{0.1375}  & \multicolumn{1}{c|}{0.1481}  \\ \cline{1-5} \cline{7-11} 
\multicolumn{1}{|c|}{$\nu_8^{(1)}$}     & \multicolumn{1}{c|}{0.0153} & \multicolumn{1}{c|}{0.0045}   & \multicolumn{1}{c|}{0.0082}  & \multicolumn{1}{c|}{0.0224}  & \multicolumn{1}{l|}{} & \multicolumn{1}{c|}{$\nu_8^{(2)}$}     & \multicolumn{1}{c|}{0.0517} & \multicolumn{1}{c|}{0.0048}   & \multicolumn{1}{c|}{0.0446}  & \multicolumn{1}{c|}{0.0567}  \\ \cline{1-5} \cline{7-11} 
\multicolumn{1}{|c|}{$\nu_9^{(1)}$}     & \multicolumn{1}{c|}{0.0142} & \multicolumn{1}{c|}{0.0043}   & \multicolumn{1}{c|}{0.0067}  & \multicolumn{1}{c|}{0.0197}  & \multicolumn{1}{l|}{} & \multicolumn{1}{c|}{$\nu_9^{(2)}$}     & \multicolumn{1}{c|}{0.0222} & \multicolumn{1}{c|}{0.0026}   & \multicolumn{1}{c|}{0.0179}  & \multicolumn{1}{c|}{0.0250}  \\ \cline{1-5} \cline{7-11} 
\multicolumn{1}{|c|}{$\nu_{10}^{(1)}$}  & \multicolumn{1}{c|}{0.0008} & \multicolumn{1}{c|}{0.0193}   & \multicolumn{1}{c|}{0.0002}  & \multicolumn{1}{c|}{0.0715}  & \multicolumn{1}{l|}{} & \multicolumn{1}{c|}{$\nu_{10}^{(2)}$}    & \multicolumn{1}{c|}{0.0003} & \multicolumn{1}{c|}{0.0018}   & \multicolumn{1}{c|}{0.0001}  & \multicolumn{1}{c|}{0.0010}  \\ \cline{1-5} \cline{7-11} 
\multicolumn{1}{|c|}{$\gamma^{(1)}$}    & \multicolumn{1}{c|}{0.0010} & \multicolumn{1}{c|}{0.0001}   & \multicolumn{1}{c|}{0.0008}  & \multicolumn{1}{c|}{0.0011}  & \multicolumn{1}{l|}{} & \multicolumn{1}{c|}{$\gamma^{(2)}$}    & \multicolumn{1}{c|}{0.0011} & \multicolumn{1}{c|}{0.0001}   & \multicolumn{1}{c|}{0.0009}  & \multicolumn{1}{c|}{0.0013}  \\ \cline{1-5} \cline{7-11} 
\multicolumn{1}{l}{}                    & \multicolumn{1}{l}{}        & \multicolumn{1}{l}{}          & \multicolumn{1}{l}{}         & \multicolumn{1}{l}{}         &                       & \multicolumn{1}{l}{}                   & \multicolumn{1}{l}{}        & \multicolumn{1}{l}{}          & \multicolumn{1}{l}{}         & \multicolumn{1}{l}{}         \\
\multicolumn{5}{c}{Systematic Parameters}                                                                                                                           &                       &                                        &                             &                               &                              &                              \\ \cline{1-5}
\multicolumn{1}{|c|}{$\alpha$}          & \multicolumn{1}{c|}{0.0073} & \multicolumn{1}{c|}{0.0013}   & \multicolumn{1}{c|}{0.0054}  & \multicolumn{1}{c|}{0.0100}  &                       &                                        &                             &                               &                              &                              \\ \cline{1-5}
\multicolumn{1}{|c|}{$\beta$}           & \multicolumn{1}{c|}{0.0027} & \multicolumn{1}{c|}{0.0003}   & \multicolumn{1}{c|}{0.0023}  & \multicolumn{1}{c|}{0.0030}  &                       & \multicolumn{1}{l}{}                   & \multicolumn{1}{l}{}        & \multicolumn{1}{l}{}          & \multicolumn{1}{l}{}         & \multicolumn{1}{l}{}         \\ \cline{1-5}
\end{tabular}
\caption{Parameter estimates based on parametric bootstrap}\label{tab:MultiTweed Results}
\end{table}

\begin{table}[h!]
\centering
\begin{tabular}{c|r|r|r|r|r|r|}
\cline{2-7}
                                    & \multicolumn{2}{c|}{Personal Auto}                        & \multicolumn{2}{c|}{Commercial Auto}                      & \multicolumn{2}{c|}{Total}                                \\ \hline
\multicolumn{1}{|c|}{Accident Year} & \multicolumn{1}{c|}{Mean} & \multicolumn{1}{c|}{SD} & \multicolumn{1}{c|}{Mean} & \multicolumn{1}{c|}{SD} & \multicolumn{1}{c|}{Mean} & \multicolumn{1}{c|}{SD} \\ \hline
\multicolumn{1}{|c|}{2}             & 149.89                    & 302.47                        & 31.79                     & 22.73                         & 181.68                    & 318.62                        \\ \hline
\multicolumn{1}{|c|}{3}             & 654.18                    & 599.86                        & 314.57                    & 75.74                         & 968.75                    & 636.11                        \\ \hline
\multicolumn{1}{|c|}{4}             & 1,572.97                  & 937.92                        & 999.35                    & 114.05                        & 2,572.32                  & 985.10                        \\ \hline
\multicolumn{1}{|c|}{5}             & 3,410.71                  & 1,364.64                      & 3,023.16                  & 167.58                        & 6,433.87                  & 1,389.11                      \\ \hline
\multicolumn{1}{|c|}{6}             & 7,946.91                  & 2,154.56                      & 7,377.44                  & 364.11                        & 15,324.35                 & 2,314.57                      \\ \hline
\multicolumn{1}{|c|}{7}             & 20,212.86                 & 3,290.60                      & 18,393.45                 & 876.04                        & 38,606.32                 & 3,738.48                      \\ \hline
\multicolumn{1}{|c|}{8}             & 39,242.16                 & 3,837.55                      & 35,019.65                 & 1,392.14                      & 74,261.82                 & 4,661.08                      \\ \hline
\multicolumn{1}{|c|}{9}             & 62,104.63                 & 3,415.46                      & 55,170.37                 & 1,646.02                      & 117,274.99                & 4,223.66                      \\ \hline
\multicolumn{1}{|c|}{10}            & 106,151.94                & 3,750.46                      & 82,571.57                 & 2,052.32                      & 188,723.51                & 4,786.30                      \\ \hline
\end{tabular}
\caption{Cumulative Outstanding Claims Reserves by Accident Year, in \$1000s}\label{tab:cumsum outstanding}
\end{table}

\section{Conclusion}

In this work, we have motivated the class of ABRMs with no closed-form PDF and proposed a novel application of the CGMM to estimate model parameters. Our methods are efficient and use moment generating functions alone. Though we primarily focus on Tweedie and Stable marginals in theory we are not limited to these distributions. We are also not bound by the form of ABRMs considered here. In the future, a more realistic model may incorporate separate scale parameters for each cell or multiple systematic components for a richer dependence structure.

We believe the results we have obtained in simulations and in an application to Schedule P data show promise for the CGMM in insurance applications. By constructing moment conditions that produce convex objective functions, we remove any need for overspecialized optimization packages or knowledge. Estimation remains generic and independent of the chosen model. That said, the art of solving ill-posed integral equations is challenging, and in the more complicated multi-LoB models further research is likely necessary.

%% file: DispTweedie.tex

\section{Dispersion Models}\label{app: EDM}

Let $Y \sim \mathcal{N}(\mu, \sigma)$ and consider the Normal distribution function:
\begin{equation}\label{eq: Normal PDF}
f_Y(y|\mu, \sigma) = \frac{1}{\sqrt{2 \pi \sigma^2}} \exp \left\lbrace -\frac{1}{2 \sigma^2} (y-\mu)^2 \right\rbrace 
\end{equation}
Notice that $(y-\mu)^2$ is a typical notion of distance. The Euclidian metric is a natural notion of distance over $\mathbb{R}$ with which to describe errors; but how to extend it to $\mathbb{R}^{+}$, $\mathbb{Z}$, $[0,\infty)$, $\mathbb{S}$ and so on? That is, we would like to make substitutions of form $(x-\mu)^2 \rightarrow d(y;\mu)$ and $\frac{1}{\sqrt{2 \pi \sigma^2}}\rightarrow a(y; \sigma^2)$ that define new distributions such as

\begin{equation}\label{rep_ex}
f_Y(y| \mu, \sigma) = a(y; \sigma^2) \exp \left\lbrace -\frac{1}{2 \sigma^2} d(y;\mu) \right\rbrace .
\end{equation}

Bent J\o rgensen used just this approach (\cite{jorgensen1987exponential} and references therein) to create \textit{exponential dispersion models} (EDMs). Naturally, $d(y;\mu)$ will be proportional to the log-likelihood of the ``$\mu$" parameter (whatever that may represent). In this way, the theory of generalized linear models can be easily extended to non-Normal, non-Euclidean settings. It is not obvious, however, that for each $d$ there will be an appropriate normalization $a$ to produce a distribution. Put another way, given $d(y;\mu)$, can we find an $a(y; \sigma^2)$ that solves the following integral equation of the first kind:
\begin{equation}\label{eq: int for DM}
1= \int a(y; \sigma^2) \exp \left\lbrace -\frac{1}{2 \sigma^2} d(y;\mu) \right\rbrace dy .
\end{equation}
In general, determining the existence and uniqueness of such a solution is hard and replete with technical issues (see Section \ref{sec: numerical CGMM}). One strategy is to start with a solution and ``invert'' this to find the distribution. Consider a distribution $a(y; \lambda)$ with known cumulant function\footnote{Known as the partition function in physics and statistical mechanics.} denoted by $\kappa(\theta)$ and defined as the logarithm of the MGF:

\begin{equation}\label{eq: cgf}
\lambda \kappa(\theta) = \log \left\lbrace \int e^{\theta z} a(y; \lambda) dy \right\rbrace .
\end{equation}
We can easily construct a distribution called the \textit{additive exponential dispersion model}, given by $Z \sim ED^*(\lambda,\theta)$):

\begin{equation}\label{eq:AEDM}
p_{Z}(z|\theta,\lambda) = \exp\left\lbrace z \theta - \lambda\kappa(\theta) \right\rbrace a(z; \lambda)  
\end{equation}
Clearly this solves (\ref{eq: int for DM}), since
$$ \int \exp\left\lbrace z \theta - \lambda \kappa(\theta) \right\rbrace a(z; \lambda) dz = e^{-\lambda \kappa(\theta)} \int e^{z\theta} a(z; \lambda) dz = 1.$$

The transformation $(Y,\mu,\sigma^2)=(\frac{Z}{\lambda}, \kappa ' (\theta), \frac{1}{\lambda})$ gives the \textit{reproductive} exponential distribution model. Note that

\begin{equation}\label{eq:add MGF}
M_{Z}(\tau)= \exp{\lbrace \lambda [\kappa (\theta +\tau)-\kappa (\theta )] \rbrace}
\end{equation}
and 
\begin{equation}\label{eq:rep MGF}
M_{Y}(\tau)= \exp{ \lbrace\lambda [\kappa (\theta +\tau/\lambda )-\kappa (\theta )] \rbrace} ,
\end{equation}
implying that
$$ \expect[Y] = \mu = \kappa'(\theta) \equiv g(\theta) \text{  and  } \var[Y]= g'(\theta).$$

The reproductive form of the pdf (given by $Y \sim ED( \mu, \sigma^2)$) is then
\begin{equation}\label{eq:REDM}
f_{Y}(y|\mu,\sigma^2) = \exp\left\lbrace -\frac{1}{2\sigma^2} \left( -2 \ell(y,\mu) \right)\right\rbrace \tilde{a}(y; \sigma^2)  
\end{equation}
where $-\ell(y,\mu) = y g^{-1}(\mu) - \kappa(g^{-1}(\mu))$ is the negative log-likelihood\footnote{Assuming the range of $\tau$ is the same as the domain of $p$; otherwise we need to add terms.} and $\tilde{a}(y; \sigma^2)  = a(y\lambda;1/\lambda )$. Having come full circle, we can regard $-2\ell$ as $d(y,\mu)$. Thus, maximizing likelihood is equivalent to minimizing the distance of some residual or error according to a metric determined by the partition function of the distribution -- a highly elegant and useful construction! Note particularly that as long as we have a distribution and corresponding MGF we can construct a model of the kind in (\ref{eq:AEDM}). Indeed, this is a very rich family of distributions found across the statistical sciences. 

\section{Tweedie Models}\label{app: Tweedie}

There is an interesting subclass of exponential dispersion models with useful properties and many well-known special cases. We begin by defining the unit variance function $V(\mu)$:

\begin{equation}\label{eq: var fun}
\var[Y] = \sigma^2 V(\mu).
\end{equation}

\noindent It can be shown that $V(\mu) = g'(g^{-1} ( \mu))$, uniquely determines the distribution. The \textit{Tweedie Exponential Dispersion models} are characterized by the unit variance function $V(\mu)=\mu^p$, or equivalently as being the only EDMs closed under scale transforms. That is, if

$$ c ED(\mu, \sigma) = ED( c\mu, c^{2-p} \sigma^2)$$

\noindent then the EDM is Tweedie (denoted $Y\sim Tw_p (\mu,\sigma^2)$). Notably, Tweedie models are also infinitely divisible (a concept discussed in the next appendix). 

The parameter $p$ in the unit variance is called the Tweedie power parameter. Solving the ODE defining the Tweedie unit variance allows us to find the $\kappa(\theta)$ that generates the Tweedie models. Let $\alpha=(p-2)/(p-1)$; then

\begin{equation}\label{eq: kp}
\kappa _{p}(\theta )={\begin{cases}{\frac {\alpha -1}{\alpha }}\left({\frac {\theta }{\alpha -1}}\right)^{\alpha },&{\text{for }}p\neq 1,2\\-\log(-\theta ),&{\text{for }}p=2\\e^{\theta },&{\text{for }}p=1.\end{cases}}
\end{equation}

Comparing this characterization to (\ref{eq:add MGF}) and (\ref{eq:rep MGF}), we can easily recover a few common distributions:

\begin{itemize}

\item $p = 0$ is the Normal distribution, 
\item $p = 1$ is the Poisson distribution,
\item $1 < p < 2$ yields the Compound Poisson–Gamma distribution, and
\item $p=2$ is the Gamma distribution.

\end{itemize} 

There are also some more exotic distributions: Inverse Gaussian ($p=3$) and some ``extreme stable" distributions ($p>3$ or $p<0$). It should be noted however that the latter distributions are only \textit{generated} by Stable distributions and will only truly be Stable if $\theta=0$. 

%% file: StableShort.tex

\begin{definition}[Stable Random Variable, 1st definition] \label{def: Stable_Def_1}

A random variable $X$ is said to have a \textit{stable distribution} if for $n\ge2$, $\exists c_n\in \mathbb{R}^+, d_n \in \mathbb{R} $ such that:

\begin{equation}\label{eq: stable_eq}
X_1 + ... + X_n \stackrel{d}{=} c_n X + d_n
\end{equation}

where the $X_i$ are independent copies of $X$.\\

\end{definition}

\noindent One can derive from this definition (\cite{zolotarev1986one}) that the characteristic function for a stable variable $X$ is given by:
\begin{equation}\label{Stable_CF}
	\phi_{X}(t) = 
	\begin{cases}
		\exp(-|\sigma t|^\alpha ( 1 - i \beta (\sign t) a ) + i t \mu ) & \alpha \ne 1 , \\
		\exp(-|\sigma t| ( 1 + i \beta \frac{2}{\pi}(\sign t) \ln |t| ) + i t \mu ) & \alpha=1 . \\
	\end{cases}
\end{equation}
Denoted $X \sim S_\alpha(\mu,\sigma,\beta)$  with $a=\tan (\frac{\pi \alpha}{2} )$. The $\mu$ and $\sigma$ are the location and scale parameters, which are equal and proportional to the mean and variance, respectively, whenever they exist. Here $\alpha \in (0,2]$ is the tail parameter. For values of $\alpha=2$ we have a Normal distribution, and for $\alpha<2$ a Pareto tailed distribution with exponent $\alpha$. The value $\beta$ is a skewness parameter and if $\alpha<1$ and $\beta = \pm 1$ support is either $[\mu,\infty)$ or $(-\infty, \mu]$. Finally, Stable random variables behave under addition operation much like the Normal. For $X_1 \sim S_\alpha(\mu_1,\sigma_1,\beta_1)$ and $X_2 \sim S_\alpha(\mu_2,\sigma_2,\beta_2)$, then $X_1 + X_2 \sim S_\alpha(\mu,\sigma,\beta)$ where:
\begin{align} 
\label{sum_prop}
\begin{split}
\mu &=\mu _{1}+\mu _{2}, \\
\sigma &=\left(\sigma_{1}^{\alpha }+\sigma_{2}^{\alpha }\right)^{\frac {1}{\alpha }}, \\
\beta &=\frac {\beta _{1}\sigma_{1}^{\alpha }+\beta _{2}\sigma_{2}^{\alpha }}{\left(\sigma_{1}^{\alpha }+\sigma_{2}^{\alpha }\right)}.
\end{split}
\end{align}

First appearing in Paul L\'{e}vy's 1925 monograph \textit{Calcul des probabilit\'{e}s}, Stable distributions went on to be studied by leading researchers, such as Andrey Kolmogorov and William Feller. One of the motivating problems of probability theory has been the distribution of sums of random variables. Stable random variables generalize the Normal as a basin of attraction to encompass all i.i.d. sums, not just the ``nice" ones with finite variance or bounded support:

\begin{theorem}{ (The generalized central limit theorem) }\label{thm: GCLT} Consider the sequence of centred and normalized sums of i.i.d RVs $ Y_{i}$ with Pareto tails such that:

\begin{center}
$1-F_{Y_i}(y) \sim k_1 y^{-\alpha}$ and $F_{Y_i}(y) \sim k_2 |y|^{-\alpha}$
\end{center}

Define: 

$$ Z_n = \frac{Y_1 + ... + Y_n }{p_n} - q_n $$

and for $\alpha \neq 1,2$\footnote{In the normal case $B_n= \sqrt{n}$. See \cite{uchaikin2011chance} for Cauchy case} set:

\begin{center}
$p_n^\alpha = \frac{2 \G{\alpha} sin(\alpha \pi /2) }{ \pi (C_1+C_2)}n$ and $q_n=\expect[Y_i]$ (if it exists, zero otherwise)
\end{center}

Then $f_{Z_n} \rightarrow f_S$ weakly where $f_S$ is a standardized stable distribution. i.e

$$ Z_n \xrightarrow[]{dist.} S_\alpha(1,\beta,0)$$

\end{theorem}

While extremely useful, Stable distributions have historically been less popular than other models. This is likely due to the fact that Stable PDFs generally do not exist in closed form. There are however three notable cases where this is not true:
\begin{itemize}
\item Normal $(\alpha=2)$;

\item Cauchy (t with d.o.f=1) $(\alpha=1, \beta=0)$;

\item L\'{e}vy $(\alpha = \frac{1}{2}, \beta=1)$.
\end{itemize}

%% file: App-F.tex
\begin{table}[h!]
\centering
\begin{tabular}{cccccccccccc}
                                    &                              &                        &                        &                        &                        &                        &                        &                        &                        &                        &                         \\ \cline{3-12} 
                                    & \multicolumn{1}{c|}{}        & \multicolumn{10}{c|}{DY}                                                                                                                                                                                                                   \\ \hline
\multicolumn{1}{|c|}{AY} & \multicolumn{1}{c|}{Premium} & \multicolumn{1}{c|}{1} & \multicolumn{1}{c|}{2} & \multicolumn{1}{c|}{3} & \multicolumn{1}{c|}{4} & \multicolumn{1}{c|}{5} & \multicolumn{1}{c|}{6} & \multicolumn{1}{c|}{7} & \multicolumn{1}{c|}{8} & \multicolumn{1}{c|}{9} & \multicolumn{1}{c|}{10} \\ \hline
\multicolumn{1}{|c|}{1}             & \multicolumn{1}{c|}{62467}   & 16864                  & 15508                  & 9341                   & 3537                   & 1853                   & 1184                   & 500                    & 308                    & 338                    & 50                      \\ \cline{1-2}
\multicolumn{1}{|c|}{2}             & \multicolumn{1}{c|}{59821}   & 14528                  & 17727                  & 8747                   & 4149                   & 2252                   & 715                    & 325                    & 261                    & 255                    &                         \\ \cline{1-2}
\multicolumn{1}{|c|}{3}             & \multicolumn{1}{c|}{62968}   & 14241                  & 13763                  & 7512                   & 5207                   & 2068                   & 1674                   & 219                    & 421                    &                        &                         \\ \cline{1-2}
\multicolumn{1}{|c|}{4}             & \multicolumn{1}{c|}{64453}   & 14765                  & 14323                  & 8426                   & 6513                   & 3144                   & 1067                   & 913                    &                        &                        &                         \\ \cline{1-2}
\multicolumn{1}{|c|}{5}             & \multicolumn{1}{c|}{71185}   & 16395                  & 17038                  & 9826                   & 6381                   & 4037                   & 1839                   &                        &                        &                        &                         \\ \cline{1-2}
\multicolumn{1}{|c|}{6}             & \multicolumn{1}{c|}{82793}   & 18136                  & 21582                  & 13415                  & 8519                   & 4583                   &                        &                        &                        &                        &                         \\ \cline{1-2}
\multicolumn{1}{|c|}{7}             & \multicolumn{1}{c|}{100826}  & 24727                  & 24037                  & 15181                  & 7105                   &                        &                        &                        &                        &                        &                         \\ \cline{1-2}
\multicolumn{1}{|c|}{8}             & \multicolumn{1}{c|}{98358}   & 24749                  & 24501                  & 11830                  &                        &                        &                        &                        &                        &                        &                         \\ \cline{1-2}
\multicolumn{1}{|c|}{9}             & \multicolumn{1}{c|}{76653}   & 23063                  & 21035                  &                        &                        &                        &                        &                        &                        &                        &                         \\ \cline{1-2}
\multicolumn{1}{|c|}{10}            & \multicolumn{1}{c|}{71326}   & 20083                  &                        &                        &                        &                        &                        &                        &                        &                        &                         \\ \cline{1-2}
\\
\end{tabular}
\caption{Auto data example of \cite{zhang2013predicting}: Personal auto line (\$1,000s) }
\label{AutoData1}
\end{table}

\begin{table}[h!]
\centering
\begin{tabular}{cccccccccccc}
                                    &                              &                        &                        &                        &                        &                        &                        &                        &                        &                        &                         \\
                                    &                              &                        &                        &                        &                        &                        &                        &                        &                        &                        &                         \\ \cline{3-12} 
                                    & \multicolumn{1}{c|}{}        & \multicolumn{10}{c|}{DY}                                                                                                                                                                                                                   \\ \hline
\multicolumn{1}{|c|}{AY} & \multicolumn{1}{c|}{Premium} & \multicolumn{1}{c|}{1} & \multicolumn{1}{c|}{2} & \multicolumn{1}{c|}{3} & \multicolumn{1}{c|}{4} & \multicolumn{1}{c|}{5} & \multicolumn{1}{c|}{6} & \multicolumn{1}{c|}{7} & \multicolumn{1}{c|}{8} & \multicolumn{1}{c|}{9} & \multicolumn{1}{c|}{10} \\ \hline
\multicolumn{1}{|c|}{1}             & \multicolumn{1}{c|}{42847}   & 5407                   & 9015                   & 4641                   & 3384                   & 1695                   & 1262                   & 1425                   & 373                    & 241                    & 6                       \\ \cline{1-2}
\multicolumn{1}{|c|}{2}             & \multicolumn{1}{c|}{38829}   & 6279                   & 8725                   & 6172                   & 4494                   & 2110                   & 919                    & 447                    & 202                    & 69                     &                         \\ \cline{1-2}
\multicolumn{1}{|c|}{3}             & \multicolumn{1}{c|}{43001}   & 7256                   & 8667                   & 4778                   & 4262                   & 2884                   & 1427                   & 889                    & 493                    &                        &                         \\ \cline{1-2}
\multicolumn{1}{|c|}{4}             & \multicolumn{1}{c|}{41840}   & 5028                   & 5317                   & 4697                   & 3795                   & 2871                   & 1100                   & 657                    &                        &                        &                         \\ \cline{1-2}
\multicolumn{1}{|c|}{5}             & \multicolumn{1}{c|}{44525}   & 5721                   & 6097                   & 6389                   & 3802                   & 4306                   & 862                    &                        &                        &                        &                         \\ \cline{1-2}
\multicolumn{1}{|c|}{6}             & \multicolumn{1}{c|}{50923}   & 7413                   & 9385                   & 7772                   & 5850                   & 3383                   &                        &                        &                        &                        &                         \\ \cline{1-2}
\multicolumn{1}{|c|}{7}             & \multicolumn{1}{c|}{56601}   & 10868                  & 12337                  & 7966                   & 8531                   &                        &                        &                        &                        &                        &                         \\ \cline{1-2}
\multicolumn{1}{|c|}{8}             & \multicolumn{1}{c|}{54609}   & 10143                  & 14193                  & 8070                   &                        &                        &                        &                        &                        &                        &                         \\ \cline{1-2}
\multicolumn{1}{|c|}{9}             & \multicolumn{1}{c|}{47204}   & 9596                   & 12235                  &                        &                        &                        &                        &                        &                        &                        &                         \\ \cline{1-2}
\multicolumn{1}{|c|}{10}            & \multicolumn{1}{c|}{42412}   & 9076                   &                        &                        &                        &                        &                        &                        &                        &                        &                         \\ \cline{1-2}  \\
\end{tabular}
\caption{Auto data example of \cite{zhang2013predicting}: Commercial auto line (\$1,000s) }
\label{AutoData2}
\end{table}